\documentclass[preprintnumbers,twocolumn,prl,aps,superscriptaddress,footinbib,amsfonts,amsmath,amssymb,showpacs]{revtex4-1}

\usepackage{amssymb, bm, amsmath, graphicx, subfigure}
\usepackage[colorlinks=true,linktocpage=true,linkcolor=blue,citecolor=blue]{hyperref}
\usepackage[usenames,dvipsnames]{color}
\usepackage{subfigure}
\usepackage{slashed}
\usepackage{multirow,array}
\usepackage[utf8]{inputenc}
\usepackage{array}
\usepackage{enumerate}
\usepackage{amsfonts}
\usepackage{dsfont}
\usepackage{mathtools}
\usepackage{mathrsfs}
\usepackage{slashed}
\usepackage{latexsym}
\usepackage{hyperref}
\usepackage{epstopdf} 
\usepackage{bbm}

\pdfoutput=1

\newcommand{\barq}{{\bar q}}

\newcommand{\bars}{{\bar s}}

\begin{document}

\title{Initial state fluctuations of QCD conserved charges in heavy-ion collisions}

\author{Mauricio Martinez}
\email[Email: ]{mmarti11@ncsu.edu}
\affiliation{North Carolina State University, Raleigh, NC 27695, USA}
\author{Matthew D. Sievert}
\email[Email: ]{msievert@illinois.edu}
\affiliation{University of Illinois at Urbana-Champaign, Urbana, IL 61801, United States}
\affiliation{Rutgers University, Piscataway, NJ 08854, USA}

\author{Douglas E. Wertepny}
\email[Email: ]{wertepny@post.bgu.ac.il}
\affiliation{Ben-Gurion University of the Negev, Beer-Sheva 84105, Israel}
\author{Jacquelyn Noronha-Hostler}
\email[Email: ]{jnorhos@illinois.edu}
\affiliation{University of Illinois at Urbana-Champaign, Urbana, IL 61801, United States}
\affiliation{Rutgers University, Piscataway, NJ 08854, USA}

\begin{abstract}
We initialize the Quantum Chromodynamic conserved charges of baryon number, strangeness, and electric charge arising from gluon splitting into quark-antiquark pairs for the initial conditions of relativistic heavy-ion collisions. A new Monte Carlo procedure that can sample from a generic energy density profile is presented, called Initial Conserved Charges in Nuclear Geometry (ICCING), based on quark and gluon multiplicities derived within the color glass condensate (CGC) effective theory. We find that while baryon number and electric charge have nearly identical geometries to the energy density profile,  the initial strangeness distribution is considerable more eccentric and is produced primarily at the hot spots corresponding to temperatures of $T\gtrsim 400$ MeV for PbPb collisions at $\sqrt{s_{NN}}=5.02$ TeV. 
\end{abstract}

\date{\today}
\maketitle


%
{\it Introduction} One of the most crucial breakthroughs in the study of heavy-ion collisions was the understanding that event-by-event fluctuating initial conditions are necessary to describe two-particle correlations \cite{Takahashi:2009na} and  in particular the triangular flow $v_3$ \cite{Alver:2010gr}.  Following this revolution, initial conditions at $\mu_B=0$ first included energy density fluctuations \cite{Schenke:2011bn}, then initial flow \cite{Gardim:2011qn,Gardim:2012yp,Gale:2012rq}, and more recently the full shear stress tensor \cite{Schenke:2019pmk} (see also \cite{Liu:2015nwa,Kurkela:2018wud}).  At lower beam energies a finite net baryon density must be initialized as well, although no single approach to this has been settled on at the moment \cite{Werner:1993uh,Shen:2017bsr,Akamatsu:2018olk,Mohs:2019iee}. These approaches at finite net baryon densities occur at the nucleonic level (i.e. they do not consider partonic structure inside the nucleons) and primarily focus only on initializing the net baryon density (with the exception of \cite{Steinheimer:2008hr}).  Important steps toward incorporating baryon stopping in a CGC picture have been made as well \cite{Itakura:2003jp, McLerran:2018avb}.

Despite the focus on finite net baryon densities, significant questions still remain at $\mu_B=0$ regarding the three QCD conserved charges of baryon number B, strangeness S, and electric charge Q.  A tension remains between light and strange particle yields \cite{Floris:2014pta,Adamczyk:2017iwn}, fluctuations \cite{Bellwied:2018tkc}, and flow harmonics \cite{Takeuchi:2015ana,Almaalol:2018gjh}, and difficulties persist describing strangeness enhancement in small systems \cite{ALICE:2017jyt} (although the core-corona approach may be an alternative \cite{Kanakubo:2019ogh}). Additionally, there appears to be charge splitting both in large and small systems \cite{Adamczyk:2013kcb,Khachatryan:2016got,Sirunyan:2017quh} but the origin of the effect is still under debate \cite{Fukushima:2008xe,Belmont:2016oqp}. It is not yet clear if these issues arise from the initial conditions or medium effects (such as \cite{Pratt:2012dz,Pratt:2015jsa,Rougemont:2017tlu,Rougemont:2015ona,Greif:2017byw,Noronha-Hostler:2019ayj,Monnai:2019hkn,Oliva:2019kin}). 

\begin{figure}[h]
\begin{center}
	\includegraphics[width=\linewidth]{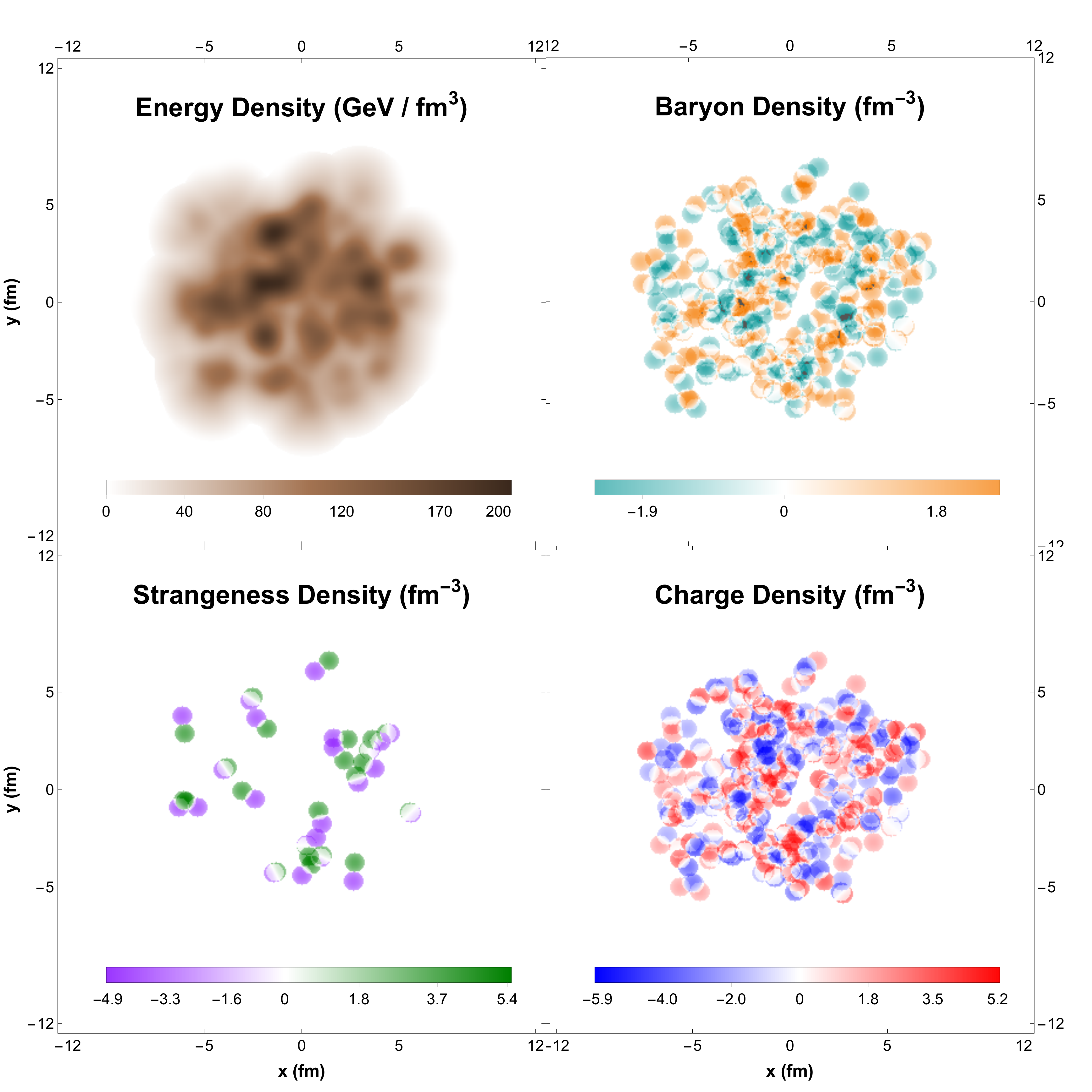}
	\caption{An event after being fully sampled by the ICCING algorithm, which supplements the initial energy density with new distributions of the three conserved charges $B, S, Q$.} 
	\label{f:Events}
	\end{center}
\end{figure}

In order to disentangle BSQ dynamics from the initial state versus the medium, we create the first 2D model of event-by-event sea quark fluctuations on top of a generic energy density profile at LHC energies as shown in Fig.~\ref{f:Events}.  Previous studies have focused on quark degrees of freedom during the approach to thermalization and chemical equilibrium \cite{Gelis:2004jp, Gelfand:2016prm, Tanji:2017xiw, Tanji:2017suk}. In this procedure we first sample gluons from a generic 2D energy density profile, then the probability of $g \rightarrow q \barq$ splitting into various flavors, and finally the displacement of the quarks relative to the gluon.  The probabilities utilized in this sampling procedure are based on multiplicities derived in a previous CGC calculation \cite{Martinez:2018ygo}. While at LHC energies the sea quarks make a subdominant contribution to the initial energy density compared to gluons \cite{Aaron:2009aa}, they provide the leading source of the conserved charges BSQ. 

%
%
%

Due to the nontrivial mass threshold of $s \bars$ pair production, the initial strangeness distribution arises not from the bulk collision geometry, but from hot spots in the initial condition generally in the region where $T\gtrsim 400$ MeV. We find that the resulting strange quark geometries are much more eccentric than the bulk, and the eccentricities of the $S^+$ and $S^-$ distributions can differ significantly from each other on an event-by-event basis.  Additionally, we find that both baryon number and electric charge experience small but finite fluctuations in their net eccentricities on an event by event basis, which can provide an important QCD contribution to measurements of charge splitting \cite{Adamczyk:2013kcb,Khachatryan:2016got,Sirunyan:2017quh}.

%

\begin{figure}[h!]
\begin{center}
	\includegraphics[width=\linewidth]{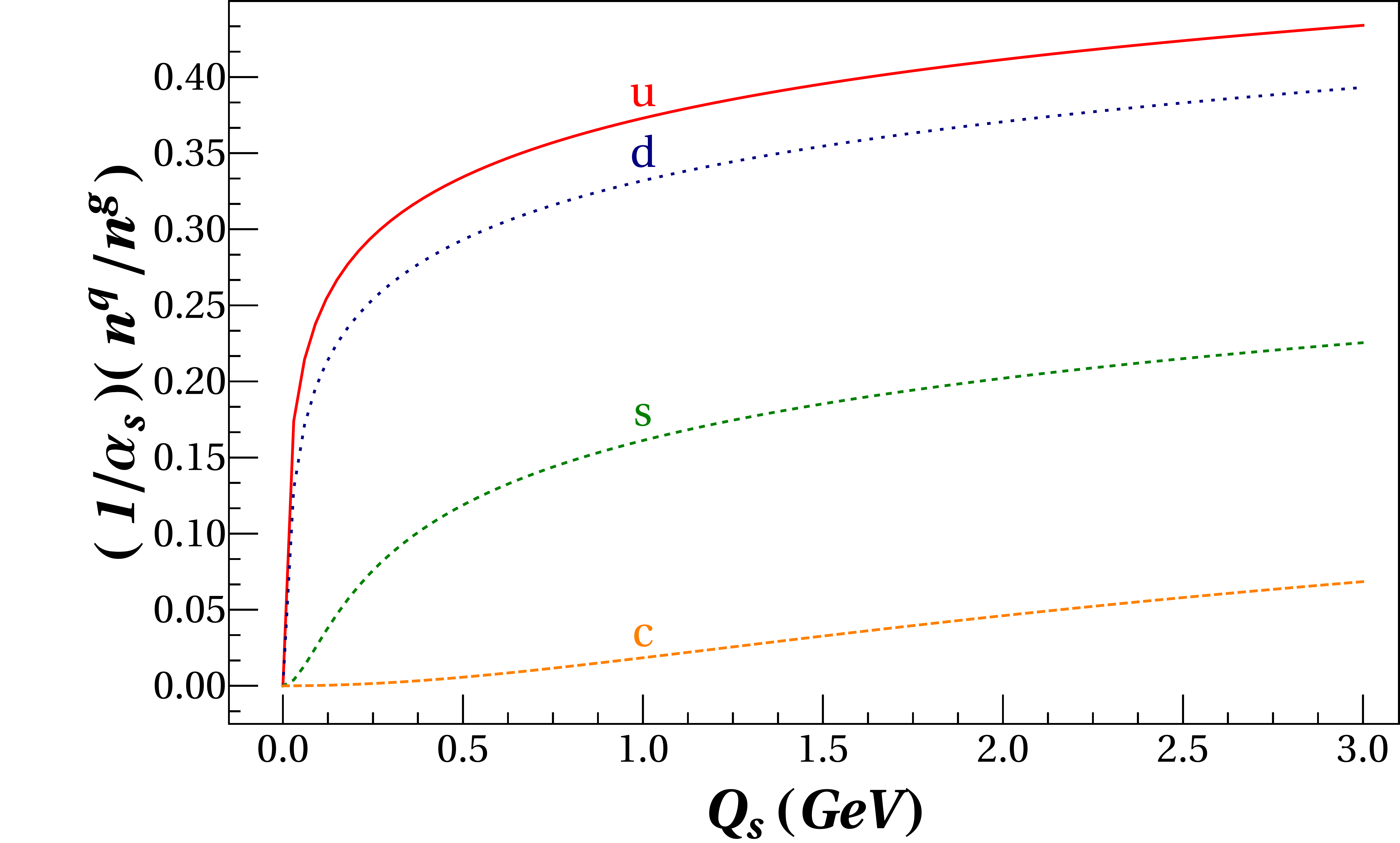}
	\caption{Quark/gluon multiplicity ratios, normalized by the coupling $\alpha_s$, as a function of the target saturation scale $Q_s$ for various quark flavors.  Here we use the McLerran-Venugopalan model~\cite{McLerran:1993ni,McLerran:1993ka} with cutoff $\Lambda / m = 0.0241$.} 
	\label{f:multratio}
	\end{center}
\end{figure}

\begin{figure}[h!]
\begin{center}
	\includegraphics[width=\linewidth]{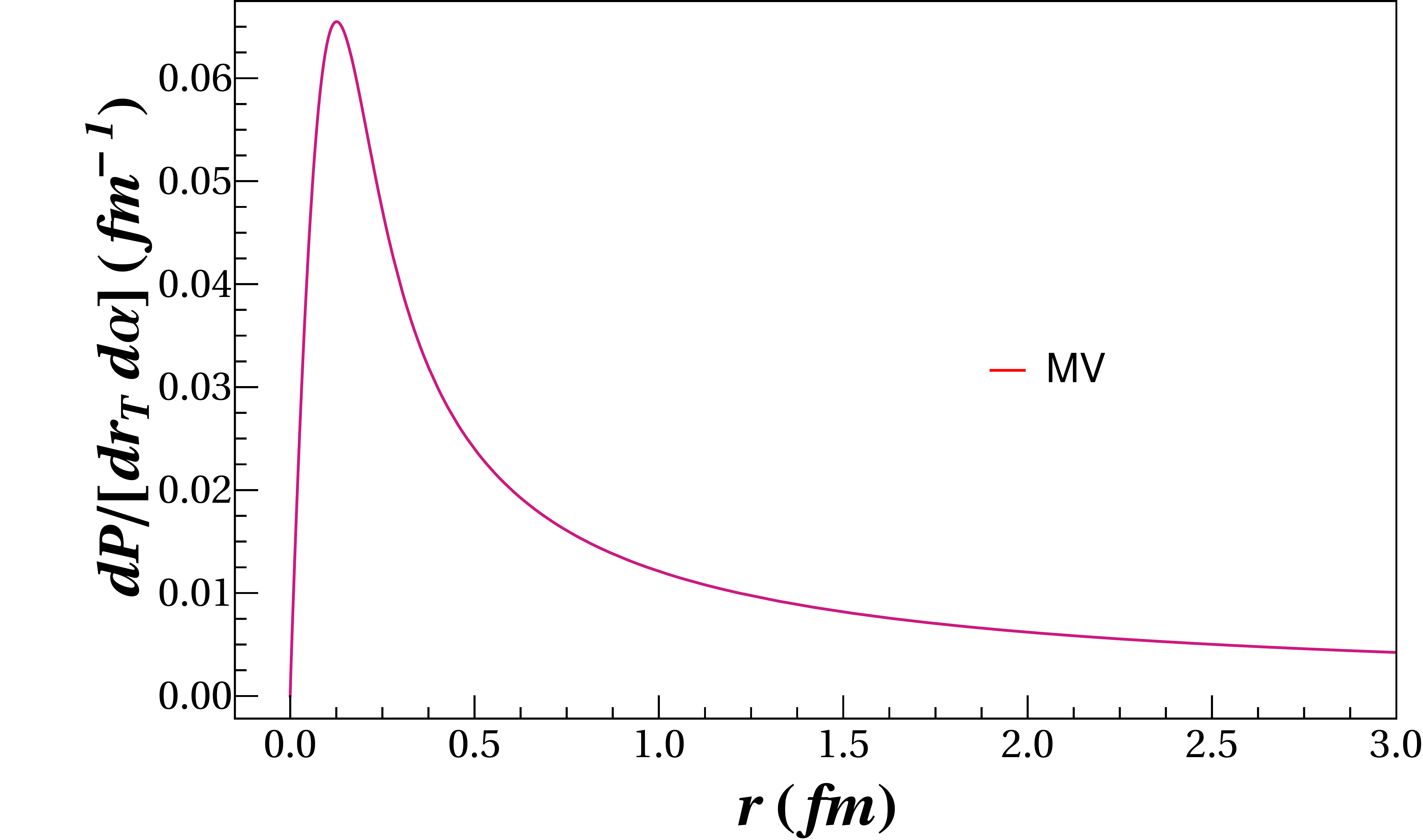}
	\caption{
	    Differential $q \barq$ splitting probability as a function of the distance $r$ between them for the  McLerran-Venugopalan (MV) model~\cite{McLerran:1993ni,McLerran:1993ka}.  In this plot we use the down quark mass $m=4.8$ MeV, representative values of the quark momentum fraction $\alpha = 0.3$ and saturation scale $Q_s=1.5$ GeV, and the cutoff in the MV model has been taken to be $\Lambda=$ 1.2 MeV.}
	\label{f:probabilities}
	\end{center}
\end{figure}

\begin{figure}[h]
\begin{center}
	\includegraphics[width=\linewidth]{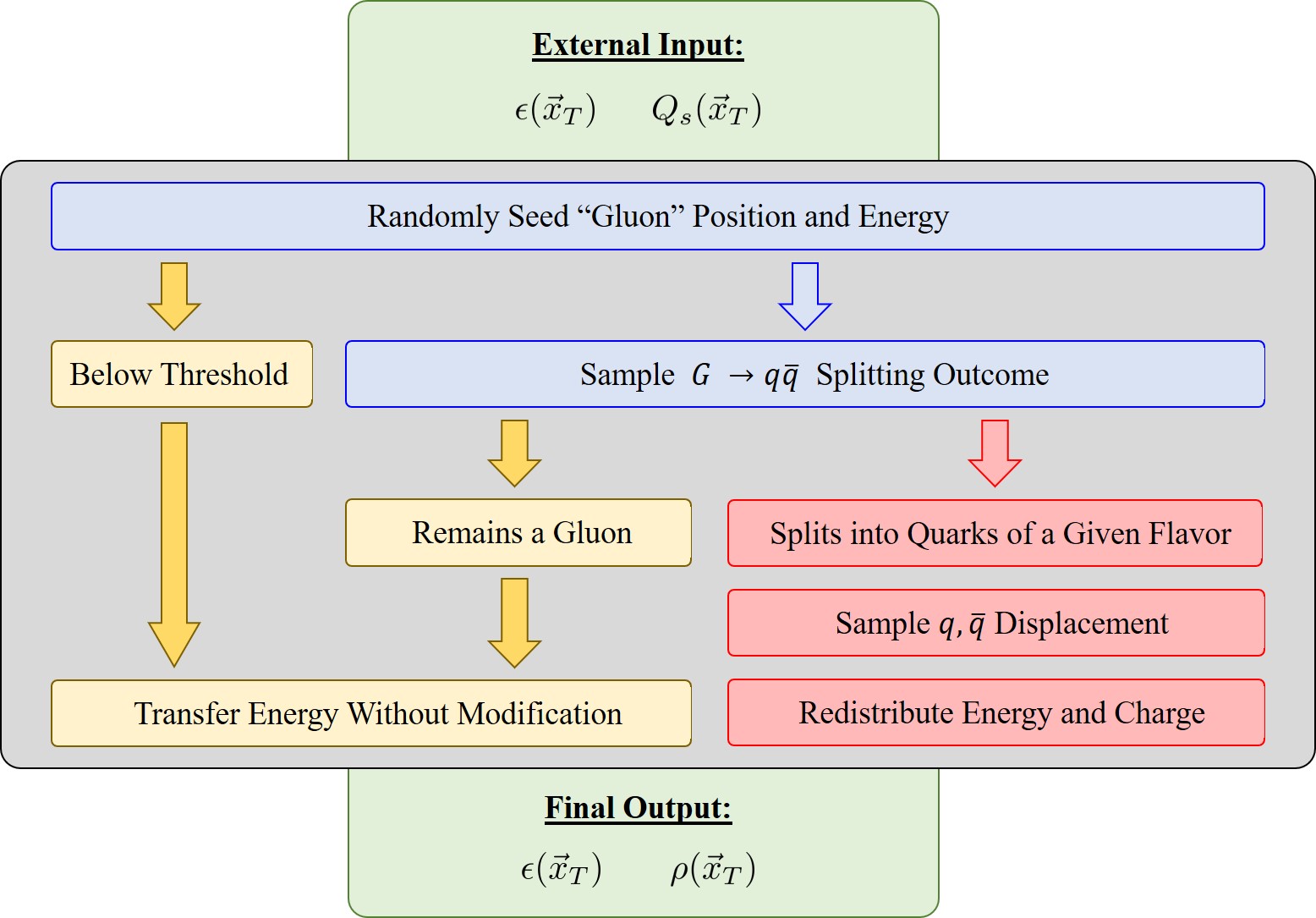}
	\caption{Flow chart of the ICCING sampling algorithm.} 
	\label{f:algorithm}
	\end{center}
\end{figure}

%
{\it The ICCING Algorithm} The model we have constructed, denoted ``Initial Conserved Charges in Nuclear Geometry (ICCING),'' performs a sampling over an initial energy density profile $\epsilon(\vec{x}_T)$ using the $g \rightarrow q \barq$ splitting functions calculated from theory to construct the corresponding sea quark distributions in the initial state.  The details of this model and its parameters are explained in the accompanying paper~\cite{longpaper}.  The theoretical ingredients used in the model are based on the calculations of \cite{Martinez:2018ygo} within the color glass condensate framework, resulting in the distributions shown in Figs.~\ref{f:multratio} and \ref{f:probabilities} for representative model parameters.  The chemistry ratios shown in Fig.~\ref{f:multratio} specify the overall probabilities to produce $q \barq$ pairs of different flavors at a given point in the transverse plane based on the saturation scale $Q_s (\vec{x}_T)$, and the probability distribution shown in Fig.~\ref{f:probabilities} specifies the $q \barq$ distance.  By performing Monte Carlo sampling of these two distributions, we take an arbitrary input energy density $\epsilon(\vec{x}_T)$ of gluons and sample it to supplement the gluon density with the accompanying distribution of quarks and antiquarks, as depicted in the flow chart of Fig.~\ref{f:algorithm}.  

%
%
%

The result is a new distribution of conserved charge densities -- baryon number $B$, strangeness $S$, and electric charge $Q$ -- along with a slightly modified energy density profile.  To demonstrate this effect we use initial energy densities derived from Trento \cite{Moreland:2014oya} for PbPb collisions at $5.02~\mathrm{TeV}$.  An example of an event after the sampling algorithm has completed is shown in Fig.~\ref{f:Events}.  The resulting picture is that, while the net total of $B$, $S$, and $Q$ is zero in the initial conditions at top collider energies, there are significant spatial fluctuations of those charge densities about zero.  These distributions couple to the event geometry in a manner which depends on the quark flavor through its mass in two ways.  The first is through the spatial variation of the chemistry ratios in Fig.~\ref{f:multratio} with the saturation scale $Q_s (\vec{x}_T)$; the second is through the mass threshold $2m$ needed to produce a quark pair of a given flavor.  The consequence of this spatial dependence is clearly visible by eye from the event shown in Fig.~\ref{f:Events}: the distributions of $B$ and $Q$ closely track the bulk geometry of the energy density, while the distribution of $S$ is qualitatively different.  Because of the nontrivial mass threshold associated with producing $s \bars$ pairs, the strangeness distribution tracks the geometry of hot spots in the event rather than the bulk, providing a different and unique probe of the initial state.  As discussed in \cite{longpaper}, this hot spot production occurs for energy densities around $\epsilon\gtrsim 50 \, \mathrm{GeV}/\mathrm{fm}^3$, which corresponds to temperatures $T\gtrsim 400$ MeV for our equation of state.

%

{\it Results}  We quantify the initial-state geometry of the bulk using the ellipticity $\varepsilon_2$ of the energy density $\epsilon$,
\begin{align} 
\varepsilon_2 \equiv \left|
\frac{\int d^2 r \, (\bm{r} - \bm{r}_{CMS})^2 \, \epsilon(\bm{r})}
{\int d^2 r \, |\bm{r} - \bm{r}_{CMS}|^2 \, \epsilon(\bm{r})} \right| ,
\end{align}
where
\begin{align}\label{eqn:com}
\bm{r}_{CMS} \equiv \frac{\int d^2 r \, \bm{r} \, \epsilon(\bm{r})}{\int d^2 r \, \epsilon(\bm{r})}
\end{align}
is the center of mass of the initial state and we use boldface to denote the complex two-vector $\bm{r} = x + i y$.  We will refer to the ellipticity associated with the energy density as the ``bulk ellipticity'' because it drives the bulk elliptic flow of all soft particles.  As we discuss in \cite{longpaper}, the standard eccentricities for a conserved charge distribution are ill-defined because the net charge is zero; hence no equivalent center of charge can be calculated analogously to the center of mass in Eq.\ \ref{eqn:com}.  Instead we quantify separately the geometry of the positive and negative regions of the conserved charge $\mathcal{X} \in \{B,S,Q\}$:
\begin{align}
\rho_{\mathcal{X}} \equiv \rho^{(\mathcal{X}^+)} \: \theta(\rho_{\mathcal{X}}) + 
\rho^{(\mathcal{X}^-)} \: \theta(-\rho_{\mathcal{X}}).
\end{align}
Then the corresponding ellipticities of positive and negative charge $\mathcal{X}$ are defined analogously:
\begin{subequations}
\begin{align}
\varepsilon_2^{(\mathcal{X}^+)} &\equiv 
\left|
\frac{
    \int d^2 r \, \left( \bm{r} - \bm{r}_{COC}^{(\mathcal{X}^+)} \right)^2 \,
    \rho^{(\mathcal{X}^+)}(\bm{r})
    }{
    \int d^2 r \, \left| \bm{r} - \bm{r}_{COC}^{(\mathcal{X}^+)} \right|^2 \,
    \rho^{(\mathcal{X}^+)}(\bm{r})
    } 
\right| ,
\\
\varepsilon_2^{(\mathcal{X}^-)} &\equiv 
\left|
\frac{
    \int d^2 r \, \left( \bm{r} - \bm{r}_{COC}^{(\mathcal{X}^-)} \right)^2 \,
    \rho^{(\mathcal{X}^-)}(\bm{r})
    }{
    \int d^2 r \, \left| \bm{r} - \bm{r}_{COC}^{(\mathcal{X}^-)} \right|^2 \,
    \rho^{(\mathcal{X}^-)}(\bm{r})
    } 
\right| ,
\end{align}
\end{subequations}
where 
\begin{align}
\bm{r}_{COC}^{(\mathcal{X}^\pm)} \equiv \frac{\int d^2 r \, \bm{r} \, \rho^{(\mathcal{X}^\pm)}(\bm{r})}{\int d^2 r \, \rho^{(\mathcal{X}^\pm)}(\bm{r})}
\end{align}
is the center of positive or negative charge.  We will also consider the two- and four-particle cumulants of the ellipticity,
\begin{subequations} \label{e:Ecumdefs}
\begin{align}   \label{e:cum2def}
\varepsilon_2 \{2\} &= \sqrt{ \left\langle \varepsilon_2^2 \right\rangle },
\\ \notag \\ \label{e:Netcumdef}
\varepsilon_2^{(\mathcal{X}, \mathrm{net})} \{2\} &= \sqrt{ \left\langle \left(
\varepsilon_2^{(\mathcal{X}^+)} - \varepsilon_2^{(\mathcal{X}^-)}
\right)^2 \right\rangle },
\\ \notag \\    \label{e:cum4def}
\varepsilon_2 \{4\} &= \sqrt[4]{
2 \left\langle \varepsilon_2^2 \right\rangle^2 -
\left\langle \varepsilon_2^4 \right\rangle
},
\end{align}
\end{subequations}
where in addition to the standard cumulants, we have also introduced in \eqref{e:Netcumdef} the second cumulant of the net ellipticity of positive versus negative charge.  While on average the ellipticity of positive charge is equal to the ellipticity of negative charge, there can be substantial fluctuations of their relative shapes as quantified by the RMS value $\varepsilon_2^{(\mathcal{X}, \mathrm{net})} \{2\}$.

%
\begin{figure}
\begin{center}
	\includegraphics[width=\linewidth]{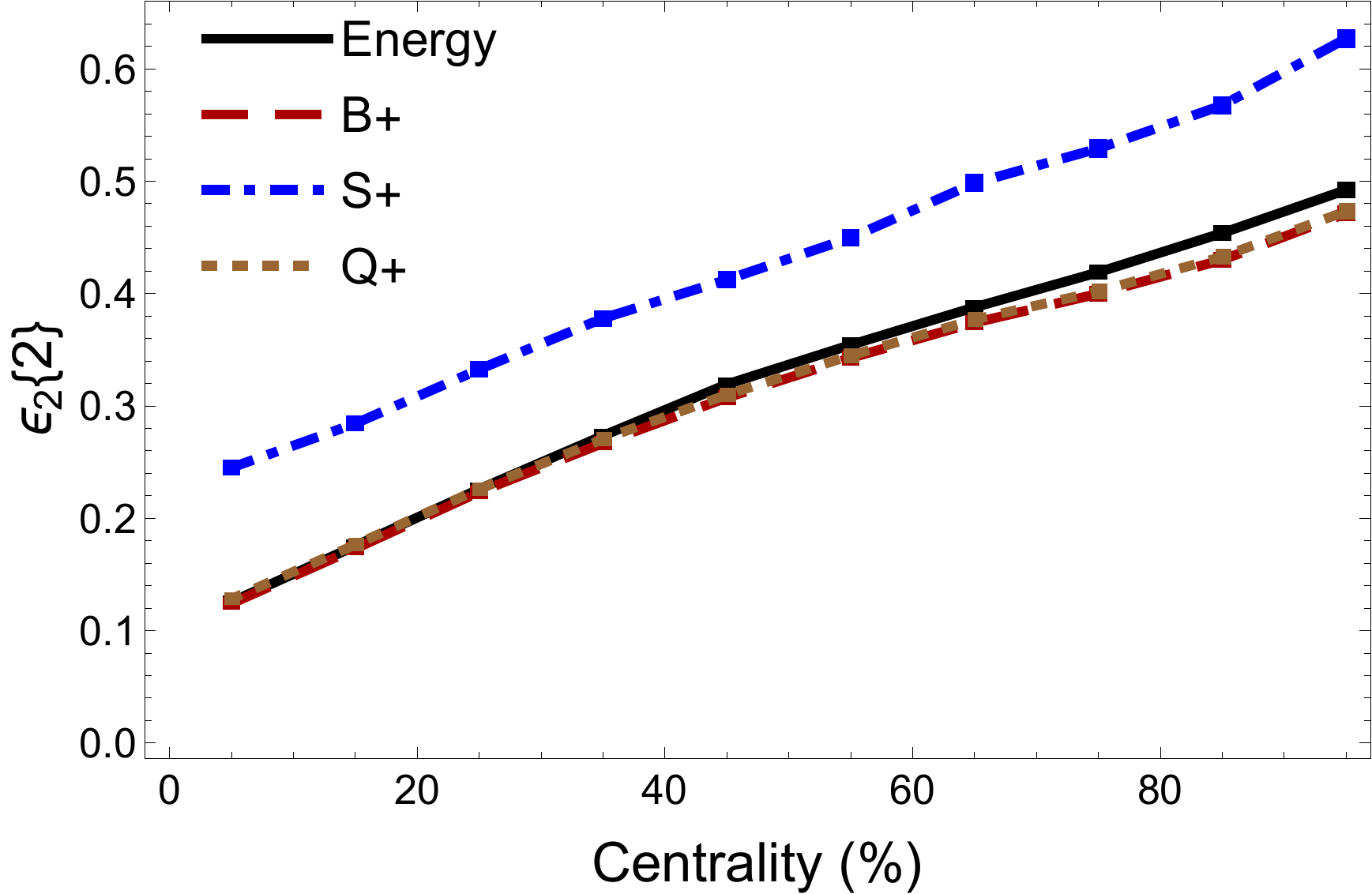}
	\caption{The two-particle cumulant $\varepsilon_2 \{2\}$ of ellipticity as a function of centrality, which measures the RMS ellipticity of the distribution.
	} 
	\label{f:Ecc2Cum2}
	\end{center}
\end{figure}
%

%
\begin{figure}
\begin{center}
	\includegraphics[width=\linewidth]{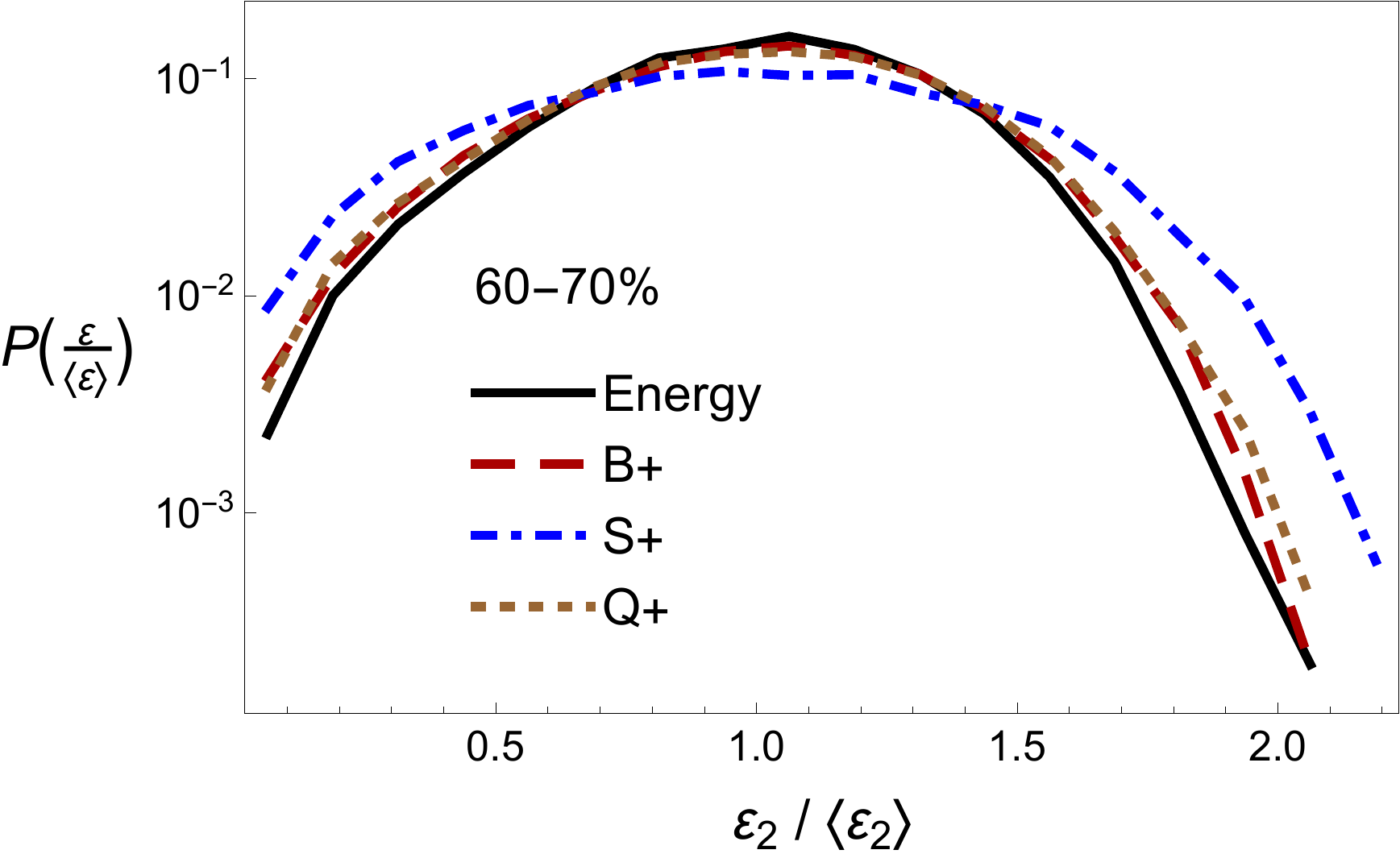}
	\caption{Histogram of the ellipticity distribution $\varepsilon_2$ of the energy density (black) as well as the regions of positive baryon number (red), strangeness (blue), and electric charge (brown).  These curves are for the 
	$60-70\%$ centrality class.} 
	\label{f:Ecc2Histo}
	\end{center}
\end{figure}
%

%
\begin{figure}
\begin{center}
	\includegraphics[width=\linewidth]{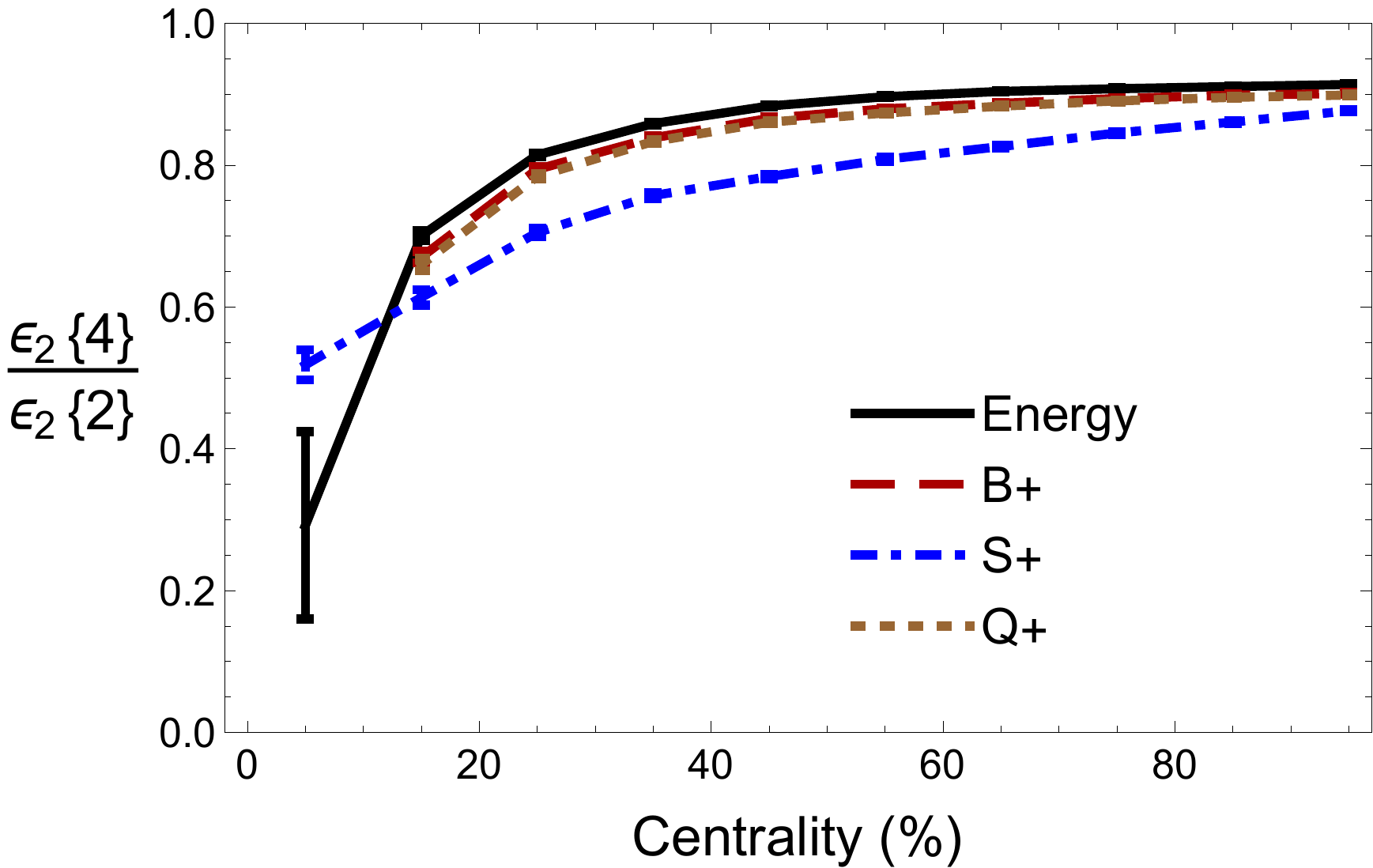}
	\caption{The ratio of the fourth and second cumulants of ellipticity as a function of centrality, which measures the event-by-event fluctuations of $\varepsilon_2$.  The closer this ratio is to unity, the fewer the fluctuations. } 
	\label{f:Ecc2Cum42}
	\end{center}
\end{figure}
%

%
\begin{figure}
\begin{center}
	\includegraphics[width=\linewidth]{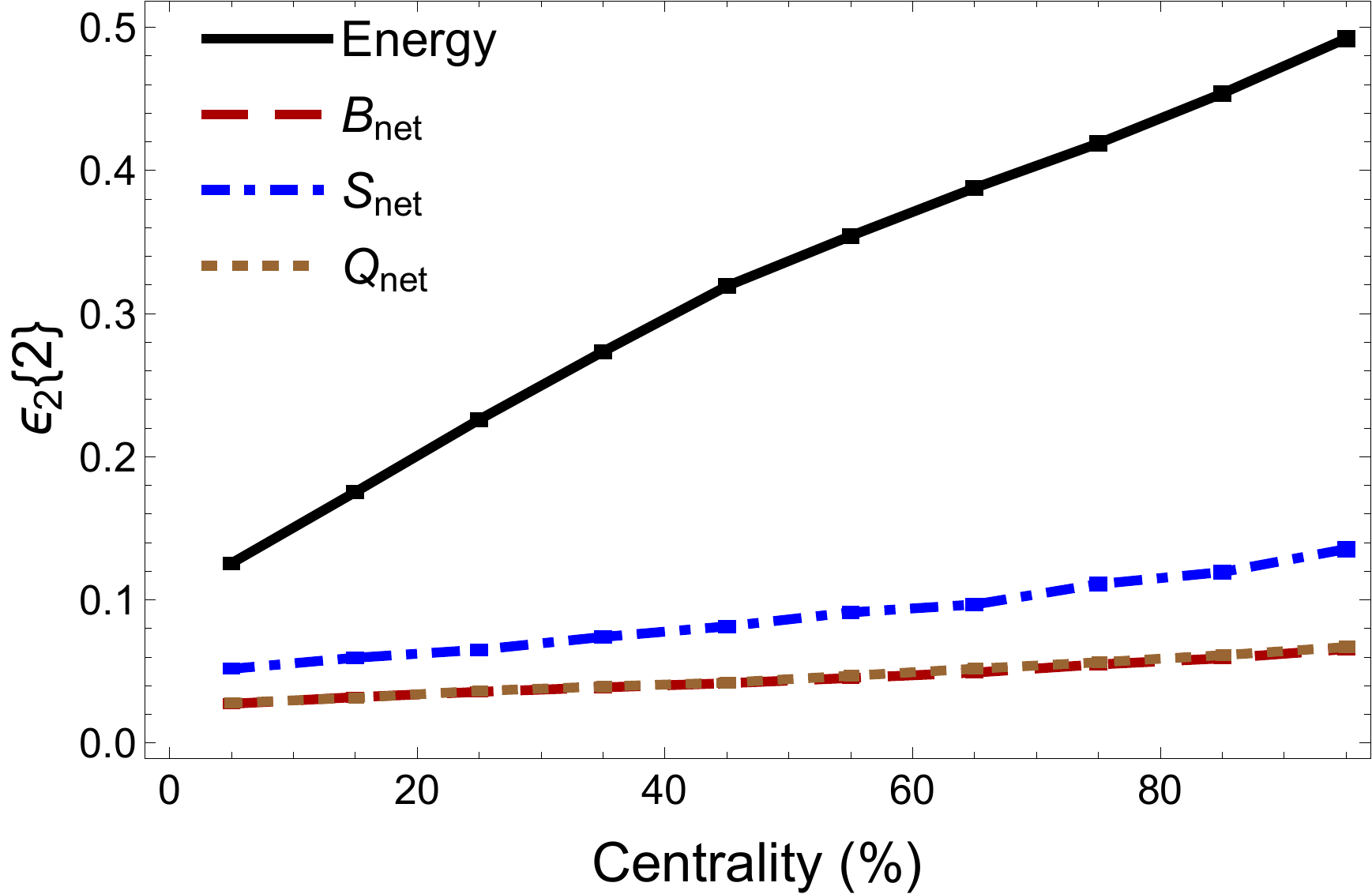}
	\caption{The two-particle cumulant of the net ellipticity \eqref{e:Netcumdef} as a function of centrality, which measures the RMS deviation in $\varepsilon_2$ between positive and negative charge.} 
	\label{f:Ecc2NetCum2}
	\end{center}
\end{figure}
%

First we consider the cumulants \eqref{e:cum2def}, \eqref{e:cum4def} of the positive-charge ellipticity; the cumulants of negative-charge ellipticity are nearly identical at LHC energies.  We note that this is foremost a way of quantifying the initial state.  Until we are able to couple these initial conditions to a full BSQ hydrodynamics code, it is unclear if the initial strangeness distribution will experience predominately linear response, as is seen for bulk charged particles \cite{Gardim:2011xv, Gardim:2014tya, Noronha-Hostler:2015dbi, Sievert:2019zjr}.  We suspect that for kaons and lambdas that only carry one strange quark, a nontrivial mixing between the bulk and strange eccentricities will occur; however, multi-strange hadrons such as cascades and omegas may be more sensitive to the initial state of strangeness.  In Fig.\ \ref{f:Ecc2Cum2} the RMS ellipticities $\varepsilon_2^{(\mathcal{X}^+)} \{2\}$ are plotted for conserved charges $\mathcal{X} \in \{B^+, S^+, Q^+\}$ compared to the bulk.  One finds that the baryon number and electric charge track the bulk energy density nearly exactly but that strangeness is significantly more eccentric. This is a natural consequence of the strange quark production that occurs in the core of an event due to fluctuating hot spots.  

The next natural question is if the distribution of the strangeness ellipticity is also affected on an event-by-event basis and, indeed, that appears to be the case, as shown in Fig.\ \ref{f:Ecc2Histo}. In Fig.\ \ref{f:Ecc2Histo} the ellipticity distributions are plotted after normalizing by the mean to remove the offsets already seen in Fig.\ \ref{f:Ecc2Cum2}.  We illustrate the centrality class $60-70\%$ because the differences are quite dramatic there and one finds a significantly wider distribution for strangeness than for the bulk.  In fact even the distributions of baryon and electric charge ellipticities are slightly wider than the bulk, but this difference is so small we do not expect it to be measurable.  

These differences in fluctuations are also reflected in the observable $\varepsilon_2 \{4\}/\varepsilon_2 \{2\}$, which is perhaps a more meaningful observable since at least for the bulk, medium effects mostly cancel out \cite{Giacalone:2017uqx,Alba:2017hhe,Sievert:2019zjr,Rao:2019vgy}.  This observable allows one to directly compare the width of the $\varepsilon_2$ distribution seen in Fig.\ \ref{f:Ecc2Histo} across centrality, with $\varepsilon_2 \{4\}/\varepsilon_2 \{2\}\rightarrow 1$ when there are fewer fluctuations.  The results plotted in Fig.~\ref{f:Ecc2Cum42} show that the fluctuations of strangeness also differ from the bulk in a systematic way.  In central collisions, the value is larger for strangeness than in bulk, reflecting fewer event-by-event fluctuations.  In mid-central and peripheral collisions, the value is smaller for strangeness, reflecting more event-by-event fluctuations.  Both of these effects indicate that the hot spot geometry of strangeness is less sensitive to the presence or absence of the overall elliptical collision geometry than the bulk.

We note that in Fig.~13 of \cite{Alba:2017hhe} it was proposed that the ratio $\varepsilon_2 \{4\}/\varepsilon_2 \{2\}$ for identified particles could be important for constraining properties of conserved charges such as BSQ diffusion \cite{Rougemont:2015ona, Rougemont:2017tlu, Greif:2017byw, Denicol:2018wdp}. In that paper no initialization of strangeness was considered and, indeed, kaons were found to track identically with pions and protons for $\varepsilon_2 \{4\}/\varepsilon_2 \{2\}$.  Depending on the strength of linear response for strangeness, it may be that $\varepsilon_2 \{4\}/\varepsilon_2 \{2\}$ of strange hadrons should, in fact, differ from light hadrons. 

Up until now, we have only considered the positive half of the conserved charge distributions.  However, it is also interesting to study quantities sensitive to the difference between the positive and negative charges, since these are subject to conservation laws.  While we reiterate that the proper definition of an eccentricity is unclear for conserved charges at $\mu_B=0$, we can use the net eccentricity defined in \eqref{e:Netcumdef} as a ``best guess" estimator from the initial state.  This quantity is a well-defined descriptor of the initial state, and it has the right symmetry properties to describe the flow of net conserved charges on an event-by-event basis.  Confirming or refuting the predictive power of this observable will require full hydrodynamic results on the final flow of strange hadrons.

In Fig.\ \ref{f:Ecc2NetCum2} we plot the second cumulant of the net ellipticities and find the rather non-trivial effect that positive and negative charge geometries can differ substantially on an event-by-event basis, even for baryon number and electric charge.  In fact, the width of the event-by-event fluctuations for B and Q is about $10\%$ of the size of the bulk ellipticity; this effect could be an important background contribution to the charge splitting of the chiral magnetic effect which needs to be taken into account.  Less surprisingly, we find once again that strangeness has larger fluctuations, in this case for the event-by-event fluctuations of the net ellipticity of strangeness.

%
{\it Conclusions}
For the first time we provide realistic event-by-event initial conditions of not only the initial energy density profile but also the initial BSQ density distributions.  These initial conditions, obtained by sampling CGC multiplicity distributions of gluons and sea quarks, indicate that strangeness is produced in the core of the collision at temperatures above $T\gtrsim 400$  MeV with a significantly more eccentric shape. 

If we find that linear response holds for strangeness, then there are a number of tantalizing consequences from this new approach.  For instance, one would then expect that strange hadrons are more likely to have a larger $v_2$ than inclusive charged particles, although care must be taken in exactly how the comparison is done.  The flow of a strange hadron such as a kaon reflects a combination of the flow seen by strange quarks and light quarks, which will partially dilute the hot spot effect seen here.  In order to avoid a further dilution, it is important to compute the $v_2 \{2\}$ from correlations of two strange hadrons with each other, rather than one strange hadron with one inclusive charged hadron.  Multi-strange hadrons, on the other hand, could see an enhancement in the flow of cascades or omegas.

Future studies will explore the initial to final state production of identified hadrons using full-fledged BSQ hydrodynamics. This may also shed light on the light vs. strange flavor hierarchy, where there is a tension between the spectra and $v_2$ in theory-to-experimental-data comparisons \cite{Takeuchi:2015ana,McDonald:2016vlt,Ryu:2017qzn,Almaalol:2018gjh}. Additionally, it may eventually be possible to constrain initial condition models including strangeness through $v_2\{4\}/v_2\{2\}$ of identified particles, analogously to what is done in the soft sector \cite{Giacalone:2017uqx,Alba:2017hhe,Sievert:2019zjr,Rao:2019vgy}.  Again, careful consideration of how to compare the flow of positive versus negative strangeness is essential.  One cannot, for example, simply subtract $v_2\{2\}$ of $K^+$ from $v_2\{2\}$ of $K^-$; this would measure the difference between the RMS flow of these particles, which we find to be zero.  Instead, one must measure $v_2\{2\}$ of the net kaon flow: the RMS of the event-by-event difference between $K^+$ and $K^-$.  We strongly encourage experimentalists to investigate this observable for different strange hadrons, which should be possible in ALICE or STAR.

Certainly these fluctuations in the geometry of positive versus negative charge are the largest for strangeness, but even baryon number and electric charge exhibit these non-trivial fluctuations.  We emphasize that with ICCING there is now a rich plethora of opportunities for studying the physics of conserved charges at the LHC simply due to local fluctuations, including the charge correlations of interest for the chiral magnetic effect in particular.  Furthermore, there is now an opportunity to constrain model parameters such as BSQ diffusion at energies with high statistics before models are then extended to the Beam Energy Scan.  Constraining these transport properties in the $\mu_B = 0$ regime, where heavy-ion collisions are best understood, is a critical baseline for studying how charge diffusion may be changed by a critical point at finite $\mu_B$.

This work also provides a path forward for combining knowledge from extracted sea quark parton distribution functions (PDFs) with initial conditions of heavy-ion collisions. In a future work we plan to make this connection explicit by replacing the CGC-motivated distributions shown in Figs.~\ref{f:multratio} and \ref{f:probabilities} with the sea quark PDFs.  Because the mass thresholds of the quarks play the largest role in the hot spot effect, we anticipate that the strange quarks will still be produced with a significantly different geometry from the bulk, but now with the overall chemistry set by the PDFs. Additionally, future studies will also explore system size effects, the inclusion of Pauli blocking between quarks \cite{Altinoluk:2016vax, Martinez:2018tuf}, pre-equilibrium effects, and varying the background energy distributions (beyond just Trento with p=0). Long term goals include the implementation of ICCING in full 3D initial conditions with nontrivial rapidity dependence and extending our sampling routine to finite baryon densities. Specifically, it would be interesting to see if the inclusion of the strange quark distribution would allow for a BSQ hydrodynamic picture to describe strangeness enhancement in small systems.

%
\acknowledgements{
{\it{Acknowledgments}}
The authors wish to thank P. Carzon, G. Denicol, A. Dumitru, J. Jalilian-Marian, Y. Kovchegov, M. Luzum, J. Nagle, P. Romatschke, T. Schaefer, V. Skokov, P. Sorensen, and P. Tribedy for useful discussions.  M.M. acknowledges support from the US-DOE Nuclear Science Grant No. DE-FG02-03ER41260 and the BEST (Beam Energy Scan Theory) DOE Topical Collaboration.  D.E.W. acknowledges support from the Zuckerman STEM Leadership Program.  J.N.H. acknowledges the support of the Alfred P. Sloan Foundation, and J.N.H. and M.D.S. acknowledge support from the US-DOE Nuclear Science Grant No. de-sc0019175, and the Office of Advanced Research Computing (OARC) at Rutgers, The State University of New Jersey for providing access to the Amarel cluster and associated research computing resources that have contributed to the results reported here.}
%

%

\begin{thebibliography}{58}%
	\makeatletter
	\providecommand \@ifxundefined [1]{%
		\@ifx{#1\undefined}
	}%
	\providecommand \@ifnum [1]{%
		\ifnum #1\expandafter \@firstoftwo
		\else \expandafter \@secondoftwo
		\fi
	}%
	\providecommand \@ifx [1]{%
		\ifx #1\expandafter \@firstoftwo
		\else \expandafter \@secondoftwo
		\fi
	}%
	\providecommand \natexlab [1]{#1}%
	\providecommand \enquote  [1]{``#1''}%
	\providecommand \bibnamefont  [1]{#1}%
	\providecommand \bibfnamefont [1]{#1}%
	\providecommand \citenamefont [1]{#1}%
	\providecommand \href@noop [0]{\@secondoftwo}%
	\providecommand \href [0]{\begingroup \@sanitize@url \@href}%
	\providecommand \@href[1]{\@@startlink{#1}\@@href}%
	\providecommand \@@href[1]{\endgroup#1\@@endlink}%
	\providecommand \@sanitize@url [0]{\catcode `\\12\catcode `\$12\catcode
		`\&12\catcode `\#12\catcode `\^12\catcode `\_12\catcode `\%12\relax}%
	\providecommand \@@startlink[1]{}%
	\providecommand \@@endlink[0]{}%
	\providecommand \url  [0]{\begingroup\@sanitize@url \@url }%
	\providecommand \@url [1]{\endgroup\@href {#1}{\urlprefix }}%
	\providecommand \urlprefix  [0]{URL }%
	\providecommand \Eprint [0]{\href }%
	\providecommand \doibase [0]{http://dx.doi.org/}%
	\providecommand \selectlanguage [0]{\@gobble}%
	\providecommand \bibinfo  [0]{\@secondoftwo}%
	\providecommand \bibfield  [0]{\@secondoftwo}%
	\providecommand \translation [1]{[#1]}%
	\providecommand \BibitemOpen [0]{}%
	\providecommand \bibitemStop [0]{}%
	\providecommand \bibitemNoStop [0]{.\EOS\space}%
	\providecommand \EOS [0]{\spacefactor3000\relax}%
	\providecommand \BibitemShut  [1]{\csname bibitem#1\endcsname}%
	\let\auto@bib@innerbib\@empty
	\bibitem [{\citenamefont {Takahashi}\ \emph {et~al.}(2009)\citenamefont
		{Takahashi} \emph {et~al.}}]{Takahashi:2009na}%
	\BibitemOpen
	\bibfield  {author} {\bibinfo {author} {\bibfnamefont {J.}~\bibnamefont
			{Takahashi}} \emph {et~al.},\ }\href {\doibase
		10.1103/PhysRevLett.103.242301} {\bibfield  {journal} {\bibinfo  {journal}
			{Phys. Rev. Lett.}\ }\textbf {\bibinfo {volume} {103}},\ \bibinfo {pages}
		{242301} (\bibinfo {year} {2009})},\ \Eprint {http://arxiv.org/abs/0902.4870}
	{arXiv:0902.4870 [nucl-th]} \BibitemShut {NoStop}%
	\bibitem [{\citenamefont {Alver}\ and\ \citenamefont
		{Roland}(2010)}]{Alver:2010gr}%
	\BibitemOpen
	\bibfield  {author} {\bibinfo {author} {\bibfnamefont {B.}~\bibnamefont
			{Alver}}\ and\ \bibinfo {author} {\bibfnamefont {G.}~\bibnamefont {Roland}},\
	}\href {\doibase 10.1103/PhysRevC.82.039903, 10.1103/PhysRevC.81.054905}
	{\bibfield  {journal} {\bibinfo  {journal} {Phys. Rev.}\ }\textbf {\bibinfo
			{volume} {C81}},\ \bibinfo {pages} {054905} (\bibinfo {year} {2010})},\
	\bibinfo {note} {[Erratum: Phys. Rev.C82,039903(2010)]},\ \Eprint
	{http://arxiv.org/abs/1003.0194} {arXiv:1003.0194 [nucl-th]} \BibitemShut
	{NoStop}%
	\bibitem [{\citenamefont {Schenke}\ \emph {et~al.}(2012)\citenamefont
		{Schenke}, \citenamefont {Jeon},\ and\ \citenamefont
		{Gale}}]{Schenke:2011bn}%
	\BibitemOpen
	\bibfield  {author} {\bibinfo {author} {\bibfnamefont {B.}~\bibnamefont
			{Schenke}}, \bibinfo {author} {\bibfnamefont {S.}~\bibnamefont {Jeon}}, \
		and\ \bibinfo {author} {\bibfnamefont {C.}~\bibnamefont {Gale}},\ }\href
	{\doibase 10.1103/PhysRevC.85.024901} {\bibfield  {journal} {\bibinfo
			{journal} {Phys. Rev.}\ }\textbf {\bibinfo {volume} {C85}},\ \bibinfo {pages}
		{024901} (\bibinfo {year} {2012})},\ \Eprint {http://arxiv.org/abs/1109.6289}
	{arXiv:1109.6289 [hep-ph]} \BibitemShut {NoStop}%
	\bibitem [{\citenamefont {Gardim}\ \emph {et~al.}(2011)\citenamefont {Gardim},
		\citenamefont {Grassi}, \citenamefont {Hama}, \citenamefont {Luzum},\ and\
		\citenamefont {Ollitrault}}]{Gardim:2011qn}%
	\BibitemOpen
	\bibfield  {author} {\bibinfo {author} {\bibfnamefont {F.~G.}\ \bibnamefont
			{Gardim}}, \bibinfo {author} {\bibfnamefont {F.}~\bibnamefont {Grassi}},
		\bibinfo {author} {\bibfnamefont {Y.}~\bibnamefont {Hama}}, \bibinfo {author}
		{\bibfnamefont {M.}~\bibnamefont {Luzum}}, \ and\ \bibinfo {author}
		{\bibfnamefont {J.-Y.}\ \bibnamefont {Ollitrault}},\ }\href {\doibase
		10.1103/PhysRevC.83.064901} {\bibfield  {journal} {\bibinfo  {journal} {Phys.
				Rev.}\ }\textbf {\bibinfo {volume} {C83}},\ \bibinfo {pages} {064901}
		(\bibinfo {year} {2011})},\ \Eprint {http://arxiv.org/abs/1103.4605}
	{arXiv:1103.4605 [nucl-th]} \BibitemShut {NoStop}%
	\bibitem [{\citenamefont {Gardim}\ \emph
		{et~al.}(2012{\natexlab{a}})\citenamefont {Gardim}, \citenamefont {Grassi},
		\citenamefont {Luzum},\ and\ \citenamefont {Ollitrault}}]{Gardim:2012yp}%
	\BibitemOpen
	\bibfield  {author} {\bibinfo {author} {\bibfnamefont {F.~G.}\ \bibnamefont
			{Gardim}}, \bibinfo {author} {\bibfnamefont {F.}~\bibnamefont {Grassi}},
		\bibinfo {author} {\bibfnamefont {M.}~\bibnamefont {Luzum}}, \ and\ \bibinfo
		{author} {\bibfnamefont {J.-Y.}\ \bibnamefont {Ollitrault}},\ }\href
	{\doibase 10.1103/PhysRevLett.109.202302} {\bibfield  {journal} {\bibinfo
			{journal} {Phys. Rev. Lett.}\ }\textbf {\bibinfo {volume} {109}},\ \bibinfo
		{pages} {202302} (\bibinfo {year} {2012}{\natexlab{a}})},\ \Eprint
	{http://arxiv.org/abs/1203.2882} {arXiv:1203.2882 [nucl-th]} \BibitemShut
	{NoStop}%
	\bibitem [{\citenamefont {Gale}\ \emph {et~al.}(2013)\citenamefont {Gale},
		\citenamefont {Jeon}, \citenamefont {Schenke}, \citenamefont {Tribedy},\ and\
		\citenamefont {Venugopalan}}]{Gale:2012rq}%
	\BibitemOpen
	\bibfield  {author} {\bibinfo {author} {\bibfnamefont {C.}~\bibnamefont
			{Gale}}, \bibinfo {author} {\bibfnamefont {S.}~\bibnamefont {Jeon}}, \bibinfo
		{author} {\bibfnamefont {B.}~\bibnamefont {Schenke}}, \bibinfo {author}
		{\bibfnamefont {P.}~\bibnamefont {Tribedy}}, \ and\ \bibinfo {author}
		{\bibfnamefont {R.}~\bibnamefont {Venugopalan}},\ }\href {\doibase
		10.1103/PhysRevLett.110.012302} {\bibfield  {journal} {\bibinfo  {journal}
			{Phys. Rev. Lett.}\ }\textbf {\bibinfo {volume} {110}},\ \bibinfo {pages}
		{012302} (\bibinfo {year} {2013})},\ \Eprint {http://arxiv.org/abs/1209.6330}
	{arXiv:1209.6330 [nucl-th]} \BibitemShut {NoStop}%
	\bibitem [{\citenamefont {Schenke}\ \emph {et~al.}(2019)\citenamefont
		{Schenke}, \citenamefont {Shen},\ and\ \citenamefont
		{Tribedy}}]{Schenke:2019pmk}%
	\BibitemOpen
	\bibfield  {author} {\bibinfo {author} {\bibfnamefont {B.}~\bibnamefont
			{Schenke}}, \bibinfo {author} {\bibfnamefont {C.}~\bibnamefont {Shen}}, \
		and\ \bibinfo {author} {\bibfnamefont {P.}~\bibnamefont {Tribedy}},\
	}\href@noop {} {\  (\bibinfo {year} {2019})},\ \Eprint
	{http://arxiv.org/abs/1908.06212} {arXiv:1908.06212 [nucl-th]} \BibitemShut
	{NoStop}%
	\bibitem [{\citenamefont {Liu}\ \emph {et~al.}(2015)\citenamefont {Liu},
		\citenamefont {Shen},\ and\ \citenamefont {Heinz}}]{Liu:2015nwa}%
	\BibitemOpen
	\bibfield  {author} {\bibinfo {author} {\bibfnamefont {J.}~\bibnamefont
			{Liu}}, \bibinfo {author} {\bibfnamefont {C.}~\bibnamefont {Shen}}, \ and\
		\bibinfo {author} {\bibfnamefont {U.}~\bibnamefont {Heinz}},\ }\href
	{\doibase 10.1103/PhysRevC.92.049904, 10.1103/PhysRevC.91.064906} {\bibfield
		{journal} {\bibinfo  {journal} {Phys. Rev.}\ }\textbf {\bibinfo {volume}
			{C91}},\ \bibinfo {pages} {064906} (\bibinfo {year} {2015})},\ \bibinfo
	{note} {[Erratum: Phys. Rev.C92,no.4,049904(2015)]},\ \Eprint
	{http://arxiv.org/abs/1504.02160} {arXiv:1504.02160 [nucl-th]} \BibitemShut
	{NoStop}%
	\bibitem [{\citenamefont {Kurkela}\ \emph {et~al.}(2019)\citenamefont
		{Kurkela}, \citenamefont {Mazeliauskas}, \citenamefont {Paquet},
		\citenamefont {Schlichting},\ and\ \citenamefont {Teaney}}]{Kurkela:2018wud}%
	\BibitemOpen
	\bibfield  {author} {\bibinfo {author} {\bibfnamefont {A.}~\bibnamefont
			{Kurkela}}, \bibinfo {author} {\bibfnamefont {A.}~\bibnamefont
			{Mazeliauskas}}, \bibinfo {author} {\bibfnamefont {J.-F.}\ \bibnamefont
			{Paquet}}, \bibinfo {author} {\bibfnamefont {S.}~\bibnamefont {Schlichting}},
		\ and\ \bibinfo {author} {\bibfnamefont {D.}~\bibnamefont {Teaney}},\ }\href
	{\doibase 10.1103/PhysRevLett.122.122302} {\bibfield  {journal} {\bibinfo
			{journal} {Phys. Rev. Lett.}\ }\textbf {\bibinfo {volume} {122}},\ \bibinfo
		{pages} {122302} (\bibinfo {year} {2019})},\ \Eprint
	{http://arxiv.org/abs/1805.01604} {arXiv:1805.01604 [hep-ph]} \BibitemShut
	{NoStop}%
	\bibitem [{\citenamefont {Werner}(1993)}]{Werner:1993uh}%
	\BibitemOpen
	\bibfield  {author} {\bibinfo {author} {\bibfnamefont {K.}~\bibnamefont
			{Werner}},\ }\href {\doibase 10.1016/0370-1573(93)90078-R} {\bibfield
		{journal} {\bibinfo  {journal} {Phys. Rept.}\ }\textbf {\bibinfo {volume}
			{232}},\ \bibinfo {pages} {87} (\bibinfo {year} {1993})}\BibitemShut
	{NoStop}%
	\bibitem [{\citenamefont {Shen}\ and\ \citenamefont
		{Schenke}(2017)}]{Shen:2017bsr}%
	\BibitemOpen
	\bibfield  {author} {\bibinfo {author} {\bibfnamefont {C.}~\bibnamefont
			{Shen}}\ and\ \bibinfo {author} {\bibfnamefont {B.}~\bibnamefont {Schenke}},\
	}\href@noop {} {\  (\bibinfo {year} {2017})},\ \Eprint
	{http://arxiv.org/abs/1710.00881} {arXiv:1710.00881 [nucl-th]} \BibitemShut
	{NoStop}%
	\bibitem [{\citenamefont {Akamatsu}\ \emph {et~al.}(2018)\citenamefont
		{Akamatsu}, \citenamefont {Asakawa}, \citenamefont {Hirano}, \citenamefont
		{Kitazawa}, \citenamefont {Morita}, \citenamefont {Murase}, \citenamefont
		{Nara}, \citenamefont {Nonaka},\ and\ \citenamefont
		{Ohnishi}}]{Akamatsu:2018olk}%
	\BibitemOpen
	\bibfield  {author} {\bibinfo {author} {\bibfnamefont {Y.}~\bibnamefont
			{Akamatsu}}, \bibinfo {author} {\bibfnamefont {M.}~\bibnamefont {Asakawa}},
		\bibinfo {author} {\bibfnamefont {T.}~\bibnamefont {Hirano}}, \bibinfo
		{author} {\bibfnamefont {M.}~\bibnamefont {Kitazawa}}, \bibinfo {author}
		{\bibfnamefont {K.}~\bibnamefont {Morita}}, \bibinfo {author} {\bibfnamefont
			{K.}~\bibnamefont {Murase}}, \bibinfo {author} {\bibfnamefont
			{Y.}~\bibnamefont {Nara}}, \bibinfo {author} {\bibfnamefont {C.}~\bibnamefont
			{Nonaka}}, \ and\ \bibinfo {author} {\bibfnamefont {A.}~\bibnamefont
			{Ohnishi}},\ }\href {\doibase 10.1103/PhysRevC.98.024909} {\bibfield
		{journal} {\bibinfo  {journal} {Phys. Rev.}\ }\textbf {\bibinfo {volume}
			{C98}},\ \bibinfo {pages} {024909} (\bibinfo {year} {2018})},\ \Eprint
	{http://arxiv.org/abs/1805.09024} {arXiv:1805.09024 [nucl-th]} \BibitemShut
	{NoStop}%
	\bibitem [{\citenamefont {Mohs}\ \emph {et~al.}(2019)\citenamefont {Mohs},
		\citenamefont {Ryu},\ and\ \citenamefont {Elfner}}]{Mohs:2019iee}%
	\BibitemOpen
	\bibfield  {author} {\bibinfo {author} {\bibfnamefont {J.}~\bibnamefont
			{Mohs}}, \bibinfo {author} {\bibfnamefont {S.}~\bibnamefont {Ryu}}, \ and\
		\bibinfo {author} {\bibfnamefont {H.}~\bibnamefont {Elfner}},\ }\href@noop {}
	{\  (\bibinfo {year} {2019})},\ \Eprint {http://arxiv.org/abs/1909.05586}
	{arXiv:1909.05586 [nucl-th]} \BibitemShut {NoStop}%
	\bibitem [{\citenamefont {Steinheimer}\ \emph {et~al.}(2009)\citenamefont
		{Steinheimer}, \citenamefont {Mitrovski}, \citenamefont {Schuster},
		\citenamefont {Petersen}, \citenamefont {Bleicher},\ and\ \citenamefont
		{Stoecker}}]{Steinheimer:2008hr}%
	\BibitemOpen
	\bibfield  {author} {\bibinfo {author} {\bibfnamefont {J.}~\bibnamefont
			{Steinheimer}}, \bibinfo {author} {\bibfnamefont {M.}~\bibnamefont
			{Mitrovski}}, \bibinfo {author} {\bibfnamefont {T.}~\bibnamefont {Schuster}},
		\bibinfo {author} {\bibfnamefont {H.}~\bibnamefont {Petersen}}, \bibinfo
		{author} {\bibfnamefont {M.}~\bibnamefont {Bleicher}}, \ and\ \bibinfo
		{author} {\bibfnamefont {H.}~\bibnamefont {Stoecker}},\ }\href {\doibase
		10.1016/j.physletb.2009.04.062} {\bibfield  {journal} {\bibinfo  {journal}
			{Phys. Lett.}\ }\textbf {\bibinfo {volume} {B676}},\ \bibinfo {pages} {126}
		(\bibinfo {year} {2009})},\ \Eprint {http://arxiv.org/abs/0811.4077}
	{arXiv:0811.4077 [hep-ph]} \BibitemShut {NoStop}%
	\bibitem [{\citenamefont {Itakura}\ \emph {et~al.}(2004)\citenamefont
		{Itakura}, \citenamefont {Kovchegov}, \citenamefont {McLerran},\ and\
		\citenamefont {Teaney}}]{Itakura:2003jp}%
	\BibitemOpen
	\bibfield  {author} {\bibinfo {author} {\bibfnamefont {K.}~\bibnamefont
			{Itakura}}, \bibinfo {author} {\bibfnamefont {Y.~V.}\ \bibnamefont
			{Kovchegov}}, \bibinfo {author} {\bibfnamefont {L.}~\bibnamefont {McLerran}},
		\ and\ \bibinfo {author} {\bibfnamefont {D.}~\bibnamefont {Teaney}},\ }\href
	{\doibase 10.1016/j.nuclphysa.2003.10.016} {\bibfield  {journal} {\bibinfo
			{journal} {Nucl. Phys.}\ }\textbf {\bibinfo {volume} {A730}},\ \bibinfo
		{pages} {160} (\bibinfo {year} {2004})},\ \Eprint
	{http://arxiv.org/abs/hep-ph/0305332} {arXiv:hep-ph/0305332} \BibitemShut
	{NoStop}%
	\bibitem [{\citenamefont {McLerran}\ \emph {et~al.}(2019)\citenamefont
		{McLerran}, \citenamefont {Schlichting},\ and\ \citenamefont
		{Sen}}]{McLerran:2018avb}%
	\BibitemOpen
	\bibfield  {author} {\bibinfo {author} {\bibfnamefont {L.~D.}\ \bibnamefont
			{McLerran}}, \bibinfo {author} {\bibfnamefont {S.}~\bibnamefont
			{Schlichting}}, \ and\ \bibinfo {author} {\bibfnamefont {S.}~\bibnamefont
			{Sen}},\ }\href {\doibase 10.1103/PhysRevD.99.074009} {\bibfield  {journal}
		{\bibinfo  {journal} {Phys. Rev.}\ }\textbf {\bibinfo {volume} {D99}},\
		\bibinfo {pages} {074009} (\bibinfo {year} {2019})},\ \Eprint
	{http://arxiv.org/abs/1811.04089} {arXiv:1811.04089 [hep-ph]} \BibitemShut
	{NoStop}%
	\bibitem [{\citenamefont {Floris}(2014)}]{Floris:2014pta}%
	\BibitemOpen
	\bibfield  {author} {\bibinfo {author} {\bibfnamefont {M.}~\bibnamefont
			{Floris}},\ }\bibfield  {booktitle} {\emph {\bibinfo {booktitle}
			{{Proceedings, 24th International Conference on Ultra-Relativistic
					Nucleus-Nucleus Collisions (Quark Matter 2014): Darmstadt, Germany, May
					19-24, 2014}}},\ }\href {\doibase 10.1016/j.nuclphysa.2014.09.002} {\bibfield
		{journal} {\bibinfo  {journal} {Nucl. Phys.}\ }\textbf {\bibinfo {volume}
			{A931}},\ \bibinfo {pages} {103} (\bibinfo {year} {2014})},\ \Eprint
	{http://arxiv.org/abs/1408.6403} {arXiv:1408.6403 [nucl-ex]} \BibitemShut
	{NoStop}%
	\bibitem [{\citenamefont {Adamczyk}\ \emph {et~al.}(2017)\citenamefont
		{Adamczyk} \emph {et~al.}}]{Adamczyk:2017iwn}%
	\BibitemOpen
	\bibfield  {author} {\bibinfo {author} {\bibfnamefont {L.}~\bibnamefont
			{Adamczyk}} \emph {et~al.} (\bibinfo {collaboration} {STAR}),\ }\href
	{\doibase 10.1103/PhysRevC.96.044904} {\bibfield  {journal} {\bibinfo
			{journal} {Phys. Rev.}\ }\textbf {\bibinfo {volume} {C96}},\ \bibinfo {pages}
		{044904} (\bibinfo {year} {2017})},\ \Eprint
	{http://arxiv.org/abs/1701.07065} {arXiv:1701.07065 [nucl-ex]} \BibitemShut
	{NoStop}%
	\bibitem [{\citenamefont {Bellwied}\ \emph {et~al.}(2019)\citenamefont
		{Bellwied}, \citenamefont {Noronha-Hostler}, \citenamefont {Parotto},
		\citenamefont {Portillo~Vazquez}, \citenamefont {Ratti},\ and\ \citenamefont
		{Stafford}}]{Bellwied:2018tkc}%
	\BibitemOpen
	\bibfield  {author} {\bibinfo {author} {\bibfnamefont {R.}~\bibnamefont
			{Bellwied}}, \bibinfo {author} {\bibfnamefont {J.}~\bibnamefont
			{Noronha-Hostler}}, \bibinfo {author} {\bibfnamefont {P.}~\bibnamefont
			{Parotto}}, \bibinfo {author} {\bibfnamefont {I.}~\bibnamefont
			{Portillo~Vazquez}}, \bibinfo {author} {\bibfnamefont {C.}~\bibnamefont
			{Ratti}}, \ and\ \bibinfo {author} {\bibfnamefont {J.~M.}\ \bibnamefont
			{Stafford}},\ }\href {\doibase 10.1103/PhysRevC.99.034912} {\bibfield
		{journal} {\bibinfo  {journal} {Phys. Rev.}\ }\textbf {\bibinfo {volume}
			{C99}},\ \bibinfo {pages} {034912} (\bibinfo {year} {2019})},\ \Eprint
	{http://arxiv.org/abs/1805.00088} {arXiv:1805.00088 [hep-ph]} \BibitemShut
	{NoStop}%
	\bibitem [{\citenamefont {Takeuchi}\ \emph {et~al.}(2015)\citenamefont
		{Takeuchi}, \citenamefont {Murase}, \citenamefont {Hirano}, \citenamefont
		{Huovinen},\ and\ \citenamefont {Nara}}]{Takeuchi:2015ana}%
	\BibitemOpen
	\bibfield  {author} {\bibinfo {author} {\bibfnamefont {S.}~\bibnamefont
			{Takeuchi}}, \bibinfo {author} {\bibfnamefont {K.}~\bibnamefont {Murase}},
		\bibinfo {author} {\bibfnamefont {T.}~\bibnamefont {Hirano}}, \bibinfo
		{author} {\bibfnamefont {P.}~\bibnamefont {Huovinen}}, \ and\ \bibinfo
		{author} {\bibfnamefont {Y.}~\bibnamefont {Nara}},\ }\href {\doibase
		10.1103/PhysRevC.92.044907} {\bibfield  {journal} {\bibinfo  {journal} {Phys.
				Rev.}\ }\textbf {\bibinfo {volume} {C92}},\ \bibinfo {pages} {044907}
		(\bibinfo {year} {2015})},\ \Eprint {http://arxiv.org/abs/1505.05961}
	{arXiv:1505.05961 [nucl-th]} \BibitemShut {NoStop}%
	\bibitem [{\citenamefont {Almaalol}\ \emph {et~al.}(2019)\citenamefont
		{Almaalol}, \citenamefont {Alqahtani},\ and\ \citenamefont
		{Strickland}}]{Almaalol:2018gjh}%
	\BibitemOpen
	\bibfield  {author} {\bibinfo {author} {\bibfnamefont {D.}~\bibnamefont
			{Almaalol}}, \bibinfo {author} {\bibfnamefont {M.}~\bibnamefont {Alqahtani}},
		\ and\ \bibinfo {author} {\bibfnamefont {M.}~\bibnamefont {Strickland}},\
	}\href {\doibase 10.1103/PhysRevC.99.044902} {\bibfield  {journal} {\bibinfo
			{journal} {Phys. Rev.}\ }\textbf {\bibinfo {volume} {C99}},\ \bibinfo {pages}
		{044902} (\bibinfo {year} {2019})},\ \Eprint
	{http://arxiv.org/abs/1807.04337} {arXiv:1807.04337 [nucl-th]} \BibitemShut
	{NoStop}%
	\bibitem [{\citenamefont {Adam}\ \emph {et~al.}(2017)\citenamefont {Adam} \emph
		{et~al.}}]{ALICE:2017jyt}%
	\BibitemOpen
	\bibfield  {author} {\bibinfo {author} {\bibfnamefont {J.}~\bibnamefont
			{Adam}} \emph {et~al.} (\bibinfo {collaboration} {ALICE}),\ }\href {\doibase
		10.1038/nphys4111} {\bibfield  {journal} {\bibinfo  {journal} {Nature Phys.}\
		}\textbf {\bibinfo {volume} {13}},\ \bibinfo {pages} {535} (\bibinfo {year}
		{2017})},\ \Eprint {http://arxiv.org/abs/1606.07424} {arXiv:1606.07424
		[nucl-ex]} \BibitemShut {NoStop}%
	\bibitem [{\citenamefont {Kanakubo}\ \emph {et~al.}(2019)\citenamefont
		{Kanakubo}, \citenamefont {Tachibana},\ and\ \citenamefont
		{Hirano}}]{Kanakubo:2019ogh}%
	\BibitemOpen
	\bibfield  {author} {\bibinfo {author} {\bibfnamefont {Y.}~\bibnamefont
			{Kanakubo}}, \bibinfo {author} {\bibfnamefont {Y.}~\bibnamefont {Tachibana}},
		\ and\ \bibinfo {author} {\bibfnamefont {T.}~\bibnamefont {Hirano}},\
	}\href@noop {} {\  (\bibinfo {year} {2019})},\ \Eprint
	{http://arxiv.org/abs/1910.10556} {arXiv:1910.10556 [nucl-th]} \BibitemShut
	{NoStop}%
	\bibitem [{\citenamefont {Adamczyk}\ \emph {et~al.}(2014)\citenamefont
		{Adamczyk} \emph {et~al.}}]{Adamczyk:2013kcb}%
	\BibitemOpen
	\bibfield  {author} {\bibinfo {author} {\bibfnamefont {L.}~\bibnamefont
			{Adamczyk}} \emph {et~al.} (\bibinfo {collaboration} {STAR}),\ }\href
	{\doibase 10.1103/PhysRevC.89.044908} {\bibfield  {journal} {\bibinfo
			{journal} {Phys. Rev.}\ }\textbf {\bibinfo {volume} {C89}},\ \bibinfo {pages}
		{044908} (\bibinfo {year} {2014})},\ \Eprint {http://arxiv.org/abs/1303.0901}
	{arXiv:1303.0901 [nucl-ex]} \BibitemShut {NoStop}%
	\bibitem [{\citenamefont {Khachatryan}\ \emph {et~al.}(2017)\citenamefont
		{Khachatryan} \emph {et~al.}}]{Khachatryan:2016got}%
	\BibitemOpen
	\bibfield  {author} {\bibinfo {author} {\bibfnamefont {V.}~\bibnamefont
			{Khachatryan}} \emph {et~al.} (\bibinfo {collaboration} {CMS}),\ }\href
	{\doibase 10.1103/PhysRevLett.118.122301} {\bibfield  {journal} {\bibinfo
			{journal} {Phys. Rev. Lett.}\ }\textbf {\bibinfo {volume} {118}},\ \bibinfo
		{pages} {122301} (\bibinfo {year} {2017})},\ \Eprint
	{http://arxiv.org/abs/1610.00263} {arXiv:1610.00263 [nucl-ex]} \BibitemShut
	{NoStop}%
	\bibitem [{\citenamefont {Sirunyan}\ \emph {et~al.}(2018)\citenamefont
		{Sirunyan} \emph {et~al.}}]{Sirunyan:2017quh}%
	\BibitemOpen
	\bibfield  {author} {\bibinfo {author} {\bibfnamefont {A.~M.}\ \bibnamefont
			{Sirunyan}} \emph {et~al.} (\bibinfo {collaboration} {CMS}),\ }\href
	{\doibase 10.1103/PhysRevC.97.044912} {\bibfield  {journal} {\bibinfo
			{journal} {Phys. Rev.}\ }\textbf {\bibinfo {volume} {C97}},\ \bibinfo {pages}
		{044912} (\bibinfo {year} {2018})},\ \Eprint
	{http://arxiv.org/abs/1708.01602} {arXiv:1708.01602 [nucl-ex]} \BibitemShut
	{NoStop}%
	\bibitem [{\citenamefont {Fukushima}\ \emph {et~al.}(2008)\citenamefont
		{Fukushima}, \citenamefont {Kharzeev},\ and\ \citenamefont
		{Warringa}}]{Fukushima:2008xe}%
	\BibitemOpen
	\bibfield  {author} {\bibinfo {author} {\bibfnamefont {K.}~\bibnamefont
			{Fukushima}}, \bibinfo {author} {\bibfnamefont {D.~E.}\ \bibnamefont
			{Kharzeev}}, \ and\ \bibinfo {author} {\bibfnamefont {H.~J.}\ \bibnamefont
			{Warringa}},\ }\href {\doibase 10.1103/PhysRevD.78.074033} {\bibfield
		{journal} {\bibinfo  {journal} {Phys. Rev.}\ }\textbf {\bibinfo {volume}
			{D78}},\ \bibinfo {pages} {074033} (\bibinfo {year} {2008})},\ \Eprint
	{http://arxiv.org/abs/0808.3382} {arXiv:0808.3382 [hep-ph]} \BibitemShut
	{NoStop}%
	\bibitem [{\citenamefont {Belmont}\ and\ \citenamefont
		{Nagle}(2017)}]{Belmont:2016oqp}%
	\BibitemOpen
	\bibfield  {author} {\bibinfo {author} {\bibfnamefont {R.}~\bibnamefont
			{Belmont}}\ and\ \bibinfo {author} {\bibfnamefont {J.~L.}\ \bibnamefont
			{Nagle}},\ }\href {\doibase 10.1103/PhysRevC.96.024901} {\bibfield  {journal}
		{\bibinfo  {journal} {Phys. Rev.}\ }\textbf {\bibinfo {volume} {C96}},\
		\bibinfo {pages} {024901} (\bibinfo {year} {2017})},\ \Eprint
	{http://arxiv.org/abs/1610.07964} {arXiv:1610.07964 [nucl-th]} \BibitemShut
	{NoStop}%
	\bibitem [{\citenamefont {Pratt}(2012)}]{Pratt:2012dz}%
	\BibitemOpen
	\bibfield  {author} {\bibinfo {author} {\bibfnamefont {S.}~\bibnamefont
			{Pratt}},\ }\href {\doibase 10.1103/PhysRevLett.108.212301} {\bibfield
		{journal} {\bibinfo  {journal} {Phys. Rev. Lett.}\ }\textbf {\bibinfo
			{volume} {108}},\ \bibinfo {pages} {212301} (\bibinfo {year} {2012})},\
	\Eprint {http://arxiv.org/abs/1203.4578} {arXiv:1203.4578 [nucl-th]}
	\BibitemShut {NoStop}%
	\bibitem [{\citenamefont {Pratt}\ \emph {et~al.}(2015)\citenamefont {Pratt},
		\citenamefont {McCormack},\ and\ \citenamefont {Ratti}}]{Pratt:2015jsa}%
	\BibitemOpen
	\bibfield  {author} {\bibinfo {author} {\bibfnamefont {S.}~\bibnamefont
			{Pratt}}, \bibinfo {author} {\bibfnamefont {W.~P.}\ \bibnamefont
			{McCormack}}, \ and\ \bibinfo {author} {\bibfnamefont {C.}~\bibnamefont
			{Ratti}},\ }\href {\doibase 10.1103/PhysRevC.92.064905} {\bibfield  {journal}
		{\bibinfo  {journal} {Phys. Rev.}\ }\textbf {\bibinfo {volume} {C92}},\
		\bibinfo {pages} {064905} (\bibinfo {year} {2015})},\ \Eprint
	{http://arxiv.org/abs/1508.07031} {arXiv:1508.07031 [nucl-th]} \BibitemShut
	{NoStop}%
	\bibitem [{\citenamefont {Rougemont}\ \emph {et~al.}(2017)\citenamefont
		{Rougemont}, \citenamefont {Critelli}, \citenamefont {Noronha-Hostler},
		\citenamefont {Noronha},\ and\ \citenamefont {Ratti}}]{Rougemont:2017tlu}%
	\BibitemOpen
	\bibfield  {author} {\bibinfo {author} {\bibfnamefont {R.}~\bibnamefont
			{Rougemont}}, \bibinfo {author} {\bibfnamefont {R.}~\bibnamefont {Critelli}},
		\bibinfo {author} {\bibfnamefont {J.}~\bibnamefont {Noronha-Hostler}},
		\bibinfo {author} {\bibfnamefont {J.}~\bibnamefont {Noronha}}, \ and\
		\bibinfo {author} {\bibfnamefont {C.}~\bibnamefont {Ratti}},\ }\href
	{\doibase 10.1103/PhysRevD.96.014032} {\bibfield  {journal} {\bibinfo
			{journal} {Phys. Rev.}\ }\textbf {\bibinfo {volume} {D96}},\ \bibinfo {pages}
		{014032} (\bibinfo {year} {2017})},\ \Eprint
	{http://arxiv.org/abs/1704.05558} {arXiv:1704.05558 [hep-ph]} \BibitemShut
	{NoStop}%
	\bibitem [{\citenamefont {Rougemont}\ \emph {et~al.}(2015)\citenamefont
		{Rougemont}, \citenamefont {Noronha},\ and\ \citenamefont
		{Noronha-Hostler}}]{Rougemont:2015ona}%
	\BibitemOpen
	\bibfield  {author} {\bibinfo {author} {\bibfnamefont {R.}~\bibnamefont
			{Rougemont}}, \bibinfo {author} {\bibfnamefont {J.}~\bibnamefont {Noronha}},
		\ and\ \bibinfo {author} {\bibfnamefont {J.}~\bibnamefont
			{Noronha-Hostler}},\ }\href {\doibase 10.1103/PhysRevLett.115.202301}
	{\bibfield  {journal} {\bibinfo  {journal} {Phys. Rev. Lett.}\ }\textbf
		{\bibinfo {volume} {115}},\ \bibinfo {pages} {202301} (\bibinfo {year}
		{2015})},\ \Eprint {http://arxiv.org/abs/1507.06972} {arXiv:1507.06972
		[hep-ph]} \BibitemShut {NoStop}%
	\bibitem [{\citenamefont {Greif}\ \emph {et~al.}(2018)\citenamefont {Greif},
		\citenamefont {Fotakis}, \citenamefont {Denicol},\ and\ \citenamefont
		{Greiner}}]{Greif:2017byw}%
	\BibitemOpen
	\bibfield  {author} {\bibinfo {author} {\bibfnamefont {M.}~\bibnamefont
			{Greif}}, \bibinfo {author} {\bibfnamefont {J.~A.}\ \bibnamefont {Fotakis}},
		\bibinfo {author} {\bibfnamefont {G.~S.}\ \bibnamefont {Denicol}}, \ and\
		\bibinfo {author} {\bibfnamefont {C.}~\bibnamefont {Greiner}},\ }\href
	{\doibase 10.1103/PhysRevLett.120.242301} {\bibfield  {journal} {\bibinfo
			{journal} {Phys. Rev. Lett.}\ }\textbf {\bibinfo {volume} {120}},\ \bibinfo
		{pages} {242301} (\bibinfo {year} {2018})},\ \Eprint
	{http://arxiv.org/abs/1711.08680} {arXiv:1711.08680 [hep-ph]} \BibitemShut
	{NoStop}%
	\bibitem [{\citenamefont {Noronha-Hostler}\ \emph {et~al.}(2019)\citenamefont
		{Noronha-Hostler}, \citenamefont {Parotto}, \citenamefont {Ratti},\ and\
		\citenamefont {Stafford}}]{Noronha-Hostler:2019ayj}%
	\BibitemOpen
	\bibfield  {author} {\bibinfo {author} {\bibfnamefont {J.}~\bibnamefont
			{Noronha-Hostler}}, \bibinfo {author} {\bibfnamefont {P.}~\bibnamefont
			{Parotto}}, \bibinfo {author} {\bibfnamefont {C.}~\bibnamefont {Ratti}}, \
		and\ \bibinfo {author} {\bibfnamefont {J.~M.}\ \bibnamefont {Stafford}},\
	}\href@noop {} {\  (\bibinfo {year} {2019})},\ \Eprint
	{http://arxiv.org/abs/1902.06723} {arXiv:1902.06723 [hep-ph]} \BibitemShut
	{NoStop}%
	\bibitem [{\citenamefont {Monnai}\ \emph {et~al.}(2019)\citenamefont {Monnai},
		\citenamefont {Schenke},\ and\ \citenamefont {Shen}}]{Monnai:2019hkn}%
	\BibitemOpen
	\bibfield  {author} {\bibinfo {author} {\bibfnamefont {A.}~\bibnamefont
			{Monnai}}, \bibinfo {author} {\bibfnamefont {B.}~\bibnamefont {Schenke}}, \
		and\ \bibinfo {author} {\bibfnamefont {C.}~\bibnamefont {Shen}},\ }\href
	{\doibase 10.1103/PhysRevC.100.024907} {\bibfield  {journal} {\bibinfo
			{journal} {Phys. Rev.}\ }\textbf {\bibinfo {volume} {C100}},\ \bibinfo
		{pages} {024907} (\bibinfo {year} {2019})},\ \Eprint
	{http://arxiv.org/abs/1902.05095} {arXiv:1902.05095 [nucl-th]} \BibitemShut
	{NoStop}%
	\bibitem [{\citenamefont {Oliva}\ \emph {et~al.}(2019)\citenamefont {Oliva},
		\citenamefont {Moreau}, \citenamefont {Voronyuk},\ and\ \citenamefont
		{Bratkovskaya}}]{Oliva:2019kin}%
	\BibitemOpen
	\bibfield  {author} {\bibinfo {author} {\bibfnamefont {L.}~\bibnamefont
			{Oliva}}, \bibinfo {author} {\bibfnamefont {P.}~\bibnamefont {Moreau}},
		\bibinfo {author} {\bibfnamefont {V.}~\bibnamefont {Voronyuk}}, \ and\
		\bibinfo {author} {\bibfnamefont {E.}~\bibnamefont {Bratkovskaya}},\
	}\href@noop {} {\  (\bibinfo {year} {2019})},\ \Eprint
	{http://arxiv.org/abs/1909.06770} {arXiv:1909.06770 [nucl-th]} \BibitemShut
	{NoStop}%
	\bibitem [{\citenamefont {Gelis}\ \emph {et~al.}(2005)\citenamefont {Gelis},
		\citenamefont {Kajantie},\ and\ \citenamefont {Lappi}}]{Gelis:2004jp}%
	\BibitemOpen
	\bibfield  {author} {\bibinfo {author} {\bibfnamefont {F.}~\bibnamefont
			{Gelis}}, \bibinfo {author} {\bibfnamefont {K.}~\bibnamefont {Kajantie}}, \
		and\ \bibinfo {author} {\bibfnamefont {T.}~\bibnamefont {Lappi}},\ }\href
	{\doibase 10.1103/PhysRevC.71.024904} {\bibfield  {journal} {\bibinfo
			{journal} {Phys. Rev.}\ }\textbf {\bibinfo {volume} {C71}},\ \bibinfo {pages}
		{024904} (\bibinfo {year} {2005})},\ \Eprint
	{http://arxiv.org/abs/hep-ph/0409058} {arXiv:hep-ph/0409058 [hep-ph]}
	\BibitemShut {NoStop}%
	\bibitem [{\citenamefont {Gelfand}\ \emph {et~al.}(2016)\citenamefont
		{Gelfand}, \citenamefont {Hebenstreit},\ and\ \citenamefont
		{Berges}}]{Gelfand:2016prm}%
	\BibitemOpen
	\bibfield  {author} {\bibinfo {author} {\bibfnamefont {D.}~\bibnamefont
			{Gelfand}}, \bibinfo {author} {\bibfnamefont {F.}~\bibnamefont
			{Hebenstreit}}, \ and\ \bibinfo {author} {\bibfnamefont {J.}~\bibnamefont
			{Berges}},\ }\href {\doibase 10.1103/PhysRevD.93.085001} {\bibfield
		{journal} {\bibinfo  {journal} {Phys. Rev.}\ }\textbf {\bibinfo {volume}
			{D93}},\ \bibinfo {pages} {085001} (\bibinfo {year} {2016})},\ \Eprint
	{http://arxiv.org/abs/1601.03576} {arXiv:1601.03576 [hep-ph]} \BibitemShut
	{NoStop}%
	\bibitem [{\citenamefont {Tanji}\ and\ \citenamefont
		{Berges}(2018)}]{Tanji:2017xiw}%
	\BibitemOpen
	\bibfield  {author} {\bibinfo {author} {\bibfnamefont {N.}~\bibnamefont
			{Tanji}}\ and\ \bibinfo {author} {\bibfnamefont {J.}~\bibnamefont {Berges}},\
	}\href {\doibase 10.1103/PhysRevD.97.034013} {\bibfield  {journal} {\bibinfo
			{journal} {Phys. Rev.}\ }\textbf {\bibinfo {volume} {D97}},\ \bibinfo {pages}
		{034013} (\bibinfo {year} {2018})},\ \Eprint
	{http://arxiv.org/abs/1711.03445} {arXiv:1711.03445 [hep-ph]} \BibitemShut
	{NoStop}%
	\bibitem [{\citenamefont {Tanji}\ and\ \citenamefont
		{Venugopalan}(2017)}]{Tanji:2017suk}%
	\BibitemOpen
	\bibfield  {author} {\bibinfo {author} {\bibfnamefont {N.}~\bibnamefont
			{Tanji}}\ and\ \bibinfo {author} {\bibfnamefont {R.}~\bibnamefont
			{Venugopalan}},\ }\href {\doibase 10.1103/PhysRevD.95.094009} {\bibfield
		{journal} {\bibinfo  {journal} {Phys. Rev.}\ }\textbf {\bibinfo {volume}
			{D95}},\ \bibinfo {pages} {094009} (\bibinfo {year} {2017})},\ \Eprint
	{http://arxiv.org/abs/1703.01372} {arXiv:1703.01372 [hep-ph]} \BibitemShut
	{NoStop}%
	\bibitem [{\citenamefont {Martinez}\ \emph {et~al.}(2018)\citenamefont
		{Martinez}, \citenamefont {Sievert},\ and\ \citenamefont
		{Wertepny}}]{Martinez:2018ygo}%
	\BibitemOpen
	\bibfield  {author} {\bibinfo {author} {\bibfnamefont {M.}~\bibnamefont
			{Martinez}}, \bibinfo {author} {\bibfnamefont {M.~D.}\ \bibnamefont
			{Sievert}}, \ and\ \bibinfo {author} {\bibfnamefont {D.~E.}\ \bibnamefont
			{Wertepny}},\ }\href {\doibase 10.1007/JHEP07(2018)003} {\bibfield  {journal}
		{\bibinfo  {journal} {JHEP}\ }\textbf {\bibinfo {volume} {07}},\ \bibinfo
		{pages} {003} (\bibinfo {year} {2018})},\ \Eprint
	{http://arxiv.org/abs/1801.08986} {arXiv:1801.08986 [hep-ph]} \BibitemShut
	{NoStop}%
	\bibitem [{\citenamefont {Aaron}\ \emph {et~al.}(2010)\citenamefont {Aaron}
		\emph {et~al.}}]{Aaron:2009aa}%
	\BibitemOpen
	\bibfield  {author} {\bibinfo {author} {\bibfnamefont {F.}~\bibnamefont
			{Aaron}} \emph {et~al.} (\bibinfo {collaboration} {H1 and ZEUS
			Collaboration}),\ }\href {\doibase 10.1007/JHEP01(2010)109} {\bibfield
		{journal} {\bibinfo  {journal} {JHEP}\ }\textbf {\bibinfo {volume} {1001}},\
		\bibinfo {pages} {109} (\bibinfo {year} {2010})},\ \Eprint
	{http://arxiv.org/abs/0911.0884} {arXiv:0911.0884 [hep-ex]} \BibitemShut
	{NoStop}%
	\bibitem [{\citenamefont {McLerran}\ and\ \citenamefont
		{Venugopalan}(1994{\natexlab{a}})}]{McLerran:1993ni}%
	\BibitemOpen
	\bibfield  {author} {\bibinfo {author} {\bibfnamefont {L.~D.}\ \bibnamefont
			{McLerran}}\ and\ \bibinfo {author} {\bibfnamefont {R.}~\bibnamefont
			{Venugopalan}},\ }\href@noop {} {\bibfield  {journal} {\bibinfo  {journal}
			{Phys. Rev.}\ }\textbf {\bibinfo {volume} {D49}},\ \bibinfo {pages} {2233}
		(\bibinfo {year} {1994}{\natexlab{a}})},\ \Eprint
	{http://arxiv.org/abs/hep-ph/9309289} {hep-ph/9309289} \BibitemShut {NoStop}%
	\bibitem [{\citenamefont {McLerran}\ and\ \citenamefont
		{Venugopalan}(1994{\natexlab{b}})}]{McLerran:1993ka}%
	\BibitemOpen
	\bibfield  {author} {\bibinfo {author} {\bibfnamefont {L.~D.}\ \bibnamefont
			{McLerran}}\ and\ \bibinfo {author} {\bibfnamefont {R.}~\bibnamefont
			{Venugopalan}},\ }\href@noop {} {\bibfield  {journal} {\bibinfo  {journal}
			{Phys. Rev.}\ }\textbf {\bibinfo {volume} {D49}},\ \bibinfo {pages} {3352}
		(\bibinfo {year} {1994}{\natexlab{b}})},\ \Eprint
	{http://arxiv.org/abs/hep-ph/9311205} {hep-ph/9311205} \BibitemShut {NoStop}%
	\bibitem [{\citenamefont {Martinez}\ \emph {et~al.}(ming)\citenamefont
		{Martinez}, \citenamefont {Sievert}, \citenamefont {Werteptny},\ and\
		\citenamefont {Noronha-Hostler}}]{longpaper}%
	\BibitemOpen
	\bibfield  {author} {\bibinfo {author} {\bibfnamefont {M.}~\bibnamefont
			{Martinez}}, \bibinfo {author} {\bibfnamefont {M.}~\bibnamefont {Sievert}},
		\bibinfo {author} {\bibfnamefont {D.~E.}\ \bibnamefont {Werteptny}}, \ and\
		\bibinfo {author} {\bibfnamefont {J.}~\bibnamefont {Noronha-Hostler}},\
	}\href@noop {} {\  (\bibinfo {year} {Forthcoming})}\BibitemShut {NoStop}%
	\bibitem [{\citenamefont {Moreland}\ \emph {et~al.}(2015)\citenamefont
		{Moreland}, \citenamefont {Bernhard},\ and\ \citenamefont
		{Bass}}]{Moreland:2014oya}%
	\BibitemOpen
	\bibfield  {author} {\bibinfo {author} {\bibfnamefont {J.~S.}\ \bibnamefont
			{Moreland}}, \bibinfo {author} {\bibfnamefont {J.~E.}\ \bibnamefont
			{Bernhard}}, \ and\ \bibinfo {author} {\bibfnamefont {S.~A.}\ \bibnamefont
			{Bass}},\ }\href {\doibase 10.1103/PhysRevC.92.011901} {\bibfield  {journal}
		{\bibinfo  {journal} {Phys. Rev.}\ }\textbf {\bibinfo {volume} {C92}},\
		\bibinfo {pages} {011901} (\bibinfo {year} {2015})},\ \Eprint
	{http://arxiv.org/abs/1412.4708} {arXiv:1412.4708 [nucl-th]} \BibitemShut
	{NoStop}%
	\bibitem [{\citenamefont {Gardim}\ \emph
		{et~al.}(2012{\natexlab{b}})\citenamefont {Gardim}, \citenamefont {Grassi},
		\citenamefont {Luzum},\ and\ \citenamefont {Ollitrault}}]{Gardim:2011xv}%
	\BibitemOpen
	\bibfield  {author} {\bibinfo {author} {\bibfnamefont {F.~G.}\ \bibnamefont
			{Gardim}}, \bibinfo {author} {\bibfnamefont {F.}~\bibnamefont {Grassi}},
		\bibinfo {author} {\bibfnamefont {M.}~\bibnamefont {Luzum}}, \ and\ \bibinfo
		{author} {\bibfnamefont {J.-Y.}\ \bibnamefont {Ollitrault}},\ }\href
	{\doibase 10.1103/PhysRevC.85.024908} {\bibfield  {journal} {\bibinfo
			{journal} {Phys. Rev.}\ }\textbf {\bibinfo {volume} {C85}},\ \bibinfo {pages}
		{024908} (\bibinfo {year} {2012}{\natexlab{b}})},\ \Eprint
	{http://arxiv.org/abs/1111.6538} {arXiv:1111.6538 [nucl-th]} \BibitemShut
	{NoStop}%
	\bibitem [{\citenamefont {Gardim}\ \emph {et~al.}(2015)\citenamefont {Gardim},
		\citenamefont {Noronha-Hostler}, \citenamefont {Luzum},\ and\ \citenamefont
		{Grassi}}]{Gardim:2014tya}%
	\BibitemOpen
	\bibfield  {author} {\bibinfo {author} {\bibfnamefont {F.~G.}\ \bibnamefont
			{Gardim}}, \bibinfo {author} {\bibfnamefont {J.}~\bibnamefont
			{Noronha-Hostler}}, \bibinfo {author} {\bibfnamefont {M.}~\bibnamefont
			{Luzum}}, \ and\ \bibinfo {author} {\bibfnamefont {F.}~\bibnamefont
			{Grassi}},\ }\href {\doibase 10.1103/PhysRevC.91.034902} {\bibfield
		{journal} {\bibinfo  {journal} {Phys. Rev.}\ }\textbf {\bibinfo {volume}
			{C91}},\ \bibinfo {pages} {034902} (\bibinfo {year} {2015})},\ \Eprint
	{http://arxiv.org/abs/1411.2574} {arXiv:1411.2574 [nucl-th]} \BibitemShut
	{NoStop}%
	\bibitem [{\citenamefont {Noronha-Hostler}\ \emph {et~al.}(2016)\citenamefont
		{Noronha-Hostler}, \citenamefont {Yan}, \citenamefont {Gardim},\ and\
		\citenamefont {Ollitrault}}]{Noronha-Hostler:2015dbi}%
	\BibitemOpen
	\bibfield  {author} {\bibinfo {author} {\bibfnamefont {J.}~\bibnamefont
			{Noronha-Hostler}}, \bibinfo {author} {\bibfnamefont {L.}~\bibnamefont
			{Yan}}, \bibinfo {author} {\bibfnamefont {F.~G.}\ \bibnamefont {Gardim}}, \
		and\ \bibinfo {author} {\bibfnamefont {J.-Y.}\ \bibnamefont {Ollitrault}},\
	}\href {\doibase 10.1103/PhysRevC.93.014909} {\bibfield  {journal} {\bibinfo
			{journal} {Phys. Rev.}\ }\textbf {\bibinfo {volume} {C93}},\ \bibinfo {pages}
		{014909} (\bibinfo {year} {2016})},\ \Eprint
	{http://arxiv.org/abs/1511.03896} {arXiv:1511.03896 [nucl-th]} \BibitemShut
	{NoStop}%
	\bibitem [{\citenamefont {Sievert}\ and\ \citenamefont
		{Noronha-Hostler}(2019)}]{Sievert:2019zjr}%
	\BibitemOpen
	\bibfield  {author} {\bibinfo {author} {\bibfnamefont {M.~D.}\ \bibnamefont
			{Sievert}}\ and\ \bibinfo {author} {\bibfnamefont {J.}~\bibnamefont
			{Noronha-Hostler}},\ }\href {\doibase 10.1103/PhysRevC.100.024904} {\bibfield
		{journal} {\bibinfo  {journal} {Phys. Rev.}\ }\textbf {\bibinfo {volume}
			{C100}},\ \bibinfo {pages} {024904} (\bibinfo {year} {2019})},\ \Eprint
	{http://arxiv.org/abs/1901.01319} {arXiv:1901.01319 [nucl-th]} \BibitemShut
	{NoStop}%
	\bibitem [{\citenamefont {Giacalone}\ \emph {et~al.}(2017)\citenamefont
		{Giacalone}, \citenamefont {Noronha-Hostler},\ and\ \citenamefont
		{Ollitrault}}]{Giacalone:2017uqx}%
	\BibitemOpen
	\bibfield  {author} {\bibinfo {author} {\bibfnamefont {G.}~\bibnamefont
			{Giacalone}}, \bibinfo {author} {\bibfnamefont {J.}~\bibnamefont
			{Noronha-Hostler}}, \ and\ \bibinfo {author} {\bibfnamefont {J.-Y.}\
			\bibnamefont {Ollitrault}},\ }\href {\doibase 10.1103/PhysRevC.95.054910}
	{\bibfield  {journal} {\bibinfo  {journal} {Phys. Rev.}\ }\textbf {\bibinfo
			{volume} {C95}},\ \bibinfo {pages} {054910} (\bibinfo {year} {2017})},\
	\Eprint {http://arxiv.org/abs/1702.01730} {arXiv:1702.01730 [nucl-th]}
	\BibitemShut {NoStop}%
	\bibitem [{\citenamefont {Alba}\ \emph {et~al.}(2018)\citenamefont {Alba},
		\citenamefont {Mantovani~Sarti}, \citenamefont {Noronha}, \citenamefont
		{Noronha-Hostler}, \citenamefont {Parotto}, \citenamefont
		{Portillo~Vazquez},\ and\ \citenamefont {Ratti}}]{Alba:2017hhe}%
	\BibitemOpen
	\bibfield  {author} {\bibinfo {author} {\bibfnamefont {P.}~\bibnamefont
			{Alba}}, \bibinfo {author} {\bibfnamefont {V.}~\bibnamefont
			{Mantovani~Sarti}}, \bibinfo {author} {\bibfnamefont {J.}~\bibnamefont
			{Noronha}}, \bibinfo {author} {\bibfnamefont {J.}~\bibnamefont
			{Noronha-Hostler}}, \bibinfo {author} {\bibfnamefont {P.}~\bibnamefont
			{Parotto}}, \bibinfo {author} {\bibfnamefont {I.}~\bibnamefont
			{Portillo~Vazquez}}, \ and\ \bibinfo {author} {\bibfnamefont
			{C.}~\bibnamefont {Ratti}},\ }\href {\doibase 10.1103/PhysRevC.98.034909}
	{\bibfield  {journal} {\bibinfo  {journal} {Phys. Rev.}\ }\textbf {\bibinfo
			{volume} {C98}},\ \bibinfo {pages} {034909} (\bibinfo {year} {2018})},\
	\Eprint {http://arxiv.org/abs/1711.05207} {arXiv:1711.05207 [nucl-th]}
	\BibitemShut {NoStop}%
	\bibitem [{\citenamefont {Rao}\ \emph {et~al.}(2019)\citenamefont {Rao},
		\citenamefont {Sievert},\ and\ \citenamefont
		{Noronha-Hostler}}]{Rao:2019vgy}%
	\BibitemOpen
	\bibfield  {author} {\bibinfo {author} {\bibfnamefont {S.}~\bibnamefont
			{Rao}}, \bibinfo {author} {\bibfnamefont {M.}~\bibnamefont {Sievert}}, \ and\
		\bibinfo {author} {\bibfnamefont {J.}~\bibnamefont {Noronha-Hostler}},\
	}\href@noop {} {\  (\bibinfo {year} {2019})},\ \Eprint
	{http://arxiv.org/abs/1910.03677} {arXiv:1910.03677 [nucl-th]} \BibitemShut
	{NoStop}%
	\bibitem [{\citenamefont {Denicol}\ \emph {et~al.}(2018)\citenamefont
		{Denicol}, \citenamefont {Gale}, \citenamefont {Jeon}, \citenamefont
		{Monnai}, \citenamefont {Schenke},\ and\ \citenamefont
		{Shen}}]{Denicol:2018wdp}%
	\BibitemOpen
	\bibfield  {author} {\bibinfo {author} {\bibfnamefont {G.~S.}\ \bibnamefont
			{Denicol}}, \bibinfo {author} {\bibfnamefont {C.}~\bibnamefont {Gale}},
		\bibinfo {author} {\bibfnamefont {S.}~\bibnamefont {Jeon}}, \bibinfo {author}
		{\bibfnamefont {A.}~\bibnamefont {Monnai}}, \bibinfo {author} {\bibfnamefont
			{B.}~\bibnamefont {Schenke}}, \ and\ \bibinfo {author} {\bibfnamefont
			{C.}~\bibnamefont {Shen}},\ }\href {\doibase 10.1103/PhysRevC.98.034916}
	{\bibfield  {journal} {\bibinfo  {journal} {Phys. Rev.}\ }\textbf {\bibinfo
			{volume} {C98}},\ \bibinfo {pages} {034916} (\bibinfo {year} {2018})},\
	\Eprint {http://arxiv.org/abs/1804.10557} {arXiv:1804.10557 [nucl-th]}
	\BibitemShut {NoStop}%
	\bibitem [{\citenamefont {McDonald}\ \emph {et~al.}(2017)\citenamefont
		{McDonald}, \citenamefont {Shen}, \citenamefont {Fillion-Gourdeau},
		\citenamefont {Jeon},\ and\ \citenamefont {Gale}}]{McDonald:2016vlt}%
	\BibitemOpen
	\bibfield  {author} {\bibinfo {author} {\bibfnamefont {S.}~\bibnamefont
			{McDonald}}, \bibinfo {author} {\bibfnamefont {C.}~\bibnamefont {Shen}},
		\bibinfo {author} {\bibfnamefont {F.}~\bibnamefont {Fillion-Gourdeau}},
		\bibinfo {author} {\bibfnamefont {S.}~\bibnamefont {Jeon}}, \ and\ \bibinfo
		{author} {\bibfnamefont {C.}~\bibnamefont {Gale}},\ }\href {\doibase
		10.1103/PhysRevC.95.064913} {\bibfield  {journal} {\bibinfo  {journal} {Phys.
				Rev.}\ }\textbf {\bibinfo {volume} {C95}},\ \bibinfo {pages} {064913}
		(\bibinfo {year} {2017})},\ \Eprint {http://arxiv.org/abs/1609.02958}
	{arXiv:1609.02958 [hep-ph]} \BibitemShut {NoStop}%
	\bibitem [{\citenamefont {Ryu}\ \emph {et~al.}(2018)\citenamefont {Ryu},
		\citenamefont {Paquet}, \citenamefont {Shen}, \citenamefont {Denicol},
		\citenamefont {Schenke}, \citenamefont {Jeon},\ and\ \citenamefont
		{Gale}}]{Ryu:2017qzn}%
	\BibitemOpen
	\bibfield  {author} {\bibinfo {author} {\bibfnamefont {S.}~\bibnamefont
			{Ryu}}, \bibinfo {author} {\bibfnamefont {J.-F.}\ \bibnamefont {Paquet}},
		\bibinfo {author} {\bibfnamefont {C.}~\bibnamefont {Shen}}, \bibinfo {author}
		{\bibfnamefont {G.}~\bibnamefont {Denicol}}, \bibinfo {author} {\bibfnamefont
			{B.}~\bibnamefont {Schenke}}, \bibinfo {author} {\bibfnamefont
			{S.}~\bibnamefont {Jeon}}, \ and\ \bibinfo {author} {\bibfnamefont
			{C.}~\bibnamefont {Gale}},\ }\href {\doibase 10.1103/PhysRevC.97.034910}
	{\bibfield  {journal} {\bibinfo  {journal} {Phys. Rev.}\ }\textbf {\bibinfo
			{volume} {C97}},\ \bibinfo {pages} {034910} (\bibinfo {year} {2018})},\
	\Eprint {http://arxiv.org/abs/1704.04216} {arXiv:1704.04216 [nucl-th]}
	\BibitemShut {NoStop}%
	\bibitem [{\citenamefont {Altinoluk}\ \emph {et~al.}(2017)\citenamefont
		{Altinoluk}, \citenamefont {Armesto}, \citenamefont {Beuf}, \citenamefont
		{Kovner},\ and\ \citenamefont {Lublinsky}}]{Altinoluk:2016vax}%
	\BibitemOpen
	\bibfield  {author} {\bibinfo {author} {\bibfnamefont {T.}~\bibnamefont
			{Altinoluk}}, \bibinfo {author} {\bibfnamefont {N.}~\bibnamefont {Armesto}},
		\bibinfo {author} {\bibfnamefont {G.}~\bibnamefont {Beuf}}, \bibinfo {author}
		{\bibfnamefont {A.}~\bibnamefont {Kovner}}, \ and\ \bibinfo {author}
		{\bibfnamefont {M.}~\bibnamefont {Lublinsky}},\ }\href {\doibase
		10.1103/PhysRevD.95.034025} {\bibfield  {journal} {\bibinfo  {journal} {Phys.
				Rev.}\ }\textbf {\bibinfo {volume} {D95}},\ \bibinfo {pages} {034025}
		(\bibinfo {year} {2017})},\ \Eprint {http://arxiv.org/abs/1610.03020}
	{arXiv:1610.03020 [hep-ph]} \BibitemShut {NoStop}%
	\bibitem [{\citenamefont {Martinez}\ \emph {et~al.}(2019)\citenamefont
		{Martinez}, \citenamefont {Sievert},\ and\ \citenamefont
		{Wertepny}}]{Martinez:2018tuf}%
	\BibitemOpen
	\bibfield  {author} {\bibinfo {author} {\bibfnamefont {M.}~\bibnamefont
			{Martinez}}, \bibinfo {author} {\bibfnamefont {M.~D.}\ \bibnamefont
			{Sievert}}, \ and\ \bibinfo {author} {\bibfnamefont {D.~E.}\ \bibnamefont
			{Wertepny}},\ }\href {\doibase 10.1007/JHEP02(2019)024} {\bibfield  {journal}
		{\bibinfo  {journal} {JHEP}\ }\textbf {\bibinfo {volume} {02}},\ \bibinfo
		{pages} {024} (\bibinfo {year} {2019})},\ \Eprint
	{http://arxiv.org/abs/1808.04896} {arXiv:1808.04896 [hep-ph]} \BibitemShut
	{NoStop}%
\end{thebibliography}
%

%

\end{document}